\documentclass[fleqn,usenatbib]{mnras}

\usepackage{newtxtext,newtxmath}
\usepackage[T1]{fontenc}
\usepackage{multirow}
\usepackage{ wasysym }
\usepackage{hyperref}
\usepackage{ulem}
\DeclareRobustCommand{\VAN}[3]{#2}
\let\VANthebibliography\thebibliography
\def\thebibliography{\DeclareRobustCommand{\VAN}[3]{##3}\VANthebibliography}

\usepackage{graphicx}	% Including figure files
\usepackage{amsmath}	% Advanced maths commands
\title[Multi-mode focal reducer MAGIC for 1-m Zeiss-1000 telescope of SAO RAS]{Small telescopes being effective: MAGIC or not?}

\author[V. L. Afanasiev et al.]{Victor L. Afanasiev$\dagger$, 
Eugene A. Malygin,$^{1}$\thanks{E-mail: male@sao.ru}
Elena S. Shablovinskaya,$^{2,1}$
Roman I. Uklein,$^{1}$
\newauthor{Vladimir R. Amirkhanyan,$^{1}$
Alexander E. Perepelitsyn$^{1}$
and Irina V. Afanasieva$^{1}$}
\\
$^{1}$Special Astrophysical Observatory of RAS, Nizhny Arkhyz, 369167, Russia\\
$^{2}$ Instituto de Estudios Astrofísicos, Facultad de Ingeniería y Ciencias, Universidad Diego Portales, Santiago, Región Metropolitana, 8370191 Chile
}

\date{Accepted 05/09/2023. Received 12/08/2023; in original form 03/05/2023}

\pubyear{2023}

\newcommand{\MAGIC}{\texttt{MAGIC}}

\begin{document}
\label{firstpage}
\pagerange{\pageref{firstpage}--\pageref{lastpage}}
\maketitle

\begin{abstract}
The paper describes the \MAGIC{} multi-mode focal reducer (Monitoring of Active Galaxies by Investigation of their Cores), commissioned on the 1-m Zeiss-1000 telescope of the Special Astrophysical Observatory of the Russian Academy of Sciences in September 2020. Three observational modes are currently realised: photometry, polarimetry, and long-slit spectroscopy.
Reducing the focal length makes it possible to obtain a sufficiently large field of view for photometry and a large slit height for spectroscopy of $\sim$12$'$, as well as a large field of view for polarimetry with a quadrupole Wollaston prism of $\sim$6$'$.4. This feature makes the complex study of extended nebulae and galaxies efficient.
The \MAGIC{} capabilities are presented in examples of observations of various astronomical objects.
The spectral mode in the range of 4000-7200 \AA\AA\ provides the spectral resolution $R \sim$ 1000; for a starlike target up to 14 mag in medium-band filters with a \textit{seeing} of 1$''$ for 20 minutes of total exposure, the photometry accuracy is better than 0.01 mag and the polarization accuracy is better than 0.6\%.
Especially for the new focal reducer, an offset guide and a position angle rotation system were implemented. The results of the modernization of the baffle system in the optical scheme of the telescope for the suppression of scattered light are also described.
\end{abstract}


%%%%%%%%%%%%%%%%%%%%%%%%%%%%%%%%%%%%%%%%%%%%%%%%%%

%%%%%%%%%%%%%%%%% BODY OF PAPER %%%%%%%%%%%%%%%%%%

\section{Introduction}

The modern level of astronomical signal registration equipment and control systems allows small telescopes to solve observational tasks that were previously available only to large instruments. The operation of meter-class telescopes is not so strictly regulated by the observation schedule, which makes them more accessible for obtaining long-term observation series. Currently, plenty of monitoring campaigns are organized at small instruments worldwide for observations of relatively bright objects variable on time-scales from hours to years, as, e.g., active galactic nuclei (AGN). 

However, many of the small Cassegrain telescopes have large focal ratios, leading to a small image scale in the focal plane. Zeiss-1000 telescope \citep[with primary mirror diameter $D$ = 1 m and focal length at the Cassegrain focus $F$ = 13.3 m,][]{Z1000} of the Special Astrophysical Observatory of the Russian Academy of Sciences (SAO RAS) also has a large focal ratio of $F/13.3$. Thus, for a pixel of the linear size $p = 13.5 ~\mu$m, the scale in the focal plane is $0''.2$/pix, providing an oversampled images within typical \textit{seeing} $\beta \approx 1''$.5 at SAO \citep{Climat}. Moreover, when extended objects, e.g., nearby Seyfert galaxies, are of particular interest, the signal-to-noise ratio (S/N) does not depend more on \textit{seeing}\footnote{For star-like objects (S/N) $\sim$ $D/\beta$.} but S/N $\sim$ $p$ $\cdot$ $D$$/F$ (obviously, this is true for optical systems not burdened by scattered light, which significantly reduces S/N). The manufacturing of a focal reducer \citep{Courtes1960, Courtes1964} naturally solves these problems. Decreasing the focal ratio from $F/13.3$ to $F/6.1$ leads to a larger scale of $0''.45$/pix meeting the demands of optimal sampling \citep[e.g.][]{Howell}. Moreover, it results in a larger field of view (FoV) important for the extended objects and also in the presence of the parallel beam allowing one to introduce dispersion elements or polarization analyzers. The latter extends the number of available observation modes for flexible reactions to weather conditions and the ability to apply diverse methods of investigation of astrophysical objects.

%Based on the productive experience in the development of multi-mode cameras based on focal reducers over the past few decades, associated with the widespread use of compact medium-sized telescopes with the Ritchey-Chr\'etien system, which have a large aberration-free FoV, but not very high aperture ratio, as well as our own positive twenty-year experience of operating focal reducers at 6-m BTA SAO RAS telescope -- we 

For these reasons, considering productive experience in the development of multi-mode cameras based on focal reducers over the past few decades [e.g. focal reducer system for 1.06-m \textit{F}/14.5 Cassegrain Zeiss-telescope of the Hoher List Observatory \citep{foc_list}, OREAD focal-reducing camera for 1.27-m \textit{F}/7.6 McGraw-Hill telescope of the Michigan-Dartmouth-MIT Observatory \citep{foc_MDM}, DFOSC for Danish 1.54-m \textit{F}/8.6 telescope at La Silla Observatory \citep{dfosc}, 
and many other devices], associated with the widespread use of compact small-sized telescopes with the Ritchey-Chr\'etien system, which have a large aberration-free FoV, but not quite fast, as well as our own positive twenty-year experience of operating focal reducers \citep[devices of the SCORPIO family, Spectral Camera with Optical Reducer for Photometrical and Interferometrical Observations,][]{sco05,sco11} at 6-m BTA telescope of SAO RAS -- we have developed a multi-mode \MAGIC{} focal reducer for the Zeiss-1000 of the SAO RAS, the parameters of which are given in Table \ref{tab:magic}. This device is aimed at a wide range of observational monitoring tasks within approaches developed at SAO RAS for the last 30 years \citep{ai1,ai2,Ukl19,M20,PolRM}, unified in the MAGIC project (Monitoring of Active Galaxies by Investigation of their Cores). Among other things, in the case of Zeiss-1000, the construction of the efficient device required additional modification of the telescope components, described in this paper.

%For these reasons, we have developed a multi-mode \MAGIC{} focal reducer for the Zeiss-1000 of the SAO RAS, \textcolor{blue}{\textbf{the parameters of which are given in Table \ref{tab:magic}}}. This device is aimed at a wide range of observational monitoring tasks within approaches developed at SAO RAS for the last 30 years \citep{ai1,ai2,Ukl19,M20,PolRM}, unified in the MAGIC project (Monitoring of Active Galaxies by Investigation of their Cores). At that, we relied on the positive twenty-year experience of operating focal reducers at 6-m BTA telescope \citep{sco05,sco11}. Nevertheless, in the case of Zeiss-1000, the construction of the efficient device required additional modification of the telescope nodes, described in this paper.

The paper structure is as follows. Section~\ref{sec:modern} describes the modernization of the optomechanical scheme of the 1-m Zeiss-1000 telescope of SAO RAS, as the installation of shielding elements, rotating platform and offset guiding system. In Section~\ref{sec3} the \texttt{MAGIC} optomechanical scheme is given together with its characteristics. Section~\ref{sec4} discusses the features of observations in the modes of photometry, polarimetry, and long-slit spectroscopy and provides examples of observations.

\section{Modernization of the optical-mechanical scheme of the telescope}
\label{sec:modern}

To increase the efficiency and accuracy of observations, we have upgraded the optomechanical scheme of the 1-m Zeiss-1000 telescope, as well as created the \texttt{MAGIC} multi-mode focal reducer. As part of the modernization of the telescope design, we introduced and changed several key components of the system:
$$
\footnotesize
{\xrightarrow[]{} \fbox{Baffles} \xrightarrow[]{} \fbox{Rotator + Guide} \xrightarrow[]{} \fbox{Calibration illumination} \xrightarrow[]{}  \fbox{\MAGIC{}} }
$$
Arrows imply the path of incoming rays.
After reflection on the primary and secondary mirrors of Zeiss-1000, the light is surrounded by baffles, then crosses the automated %\textit{modernized} 
turntable consisting of the rotator and the offset guide, after which it passes through the calibration illumination module and only then enters the \texttt{MAGIC} entrance pupil. 

The modified components in the scheme complement the \texttt{MAGIC} device, however, they are permanent modifications of the entire optical system of the telescope and also work in conjunction with other devices operating at the Cassegrain focus. Nevertheless, all these modules are independent and separated from each other and can be removed if necessary. At the moment, the rotation and guiding modules are implemented on the telescope and are at the stage of the test observations. Further in the section, we will sequentially describe these components in brief detail. Being the essential part of the telescope modernization, the module of telecentric illumination with discrete and continuous spectrum sources for spectral calibrations designed similarly to the concepts implemented for the 6-m BTA telescope adapter \citep{bta_adapter}  is in the process of development and is a point of the upcoming paper.

\begin{table}
\caption{The main parameters of \MAGIC{} with a CCD on the Zeiss-1000 telescope of the SAO RAS}
\begin{tabular}{cc}
\hline
\multicolumn{2}{c}{\MAGIC{} main parameters}                                                                               \\ \hline
Input focal ratio of focal reducer                        & $F$/12.5                                                     \\
Total focal ratio at the Zeiss-1000                    & $F$/6.1                                                      \\
QE (optics + telescope + CCD)                              & $\sim$50\%                                                 \\
Image quality (FWHM)                                       & 0$''$.3                                                      \\
Spectral range                                             & 340-990 nm                                                 \\
Weight                                                     & 23 kg                                                      \\
Dimensions                                                 & 430 $\times$ 440 $\times$ 265 mm \\ \hline
CCD system                                                 & Andor iKonL-936                                            \\
CCD                                                        & E2V CCD42-40 (BEX2-DD)                                     \\
Format                                                     & 2048 $\times$ 2048 pix                         \\
Pixel size                                                 & 13.5 $\times$ 13.5 $\mu$m       \\
\multirow{2}{*}{QE}                                        & 400-850 nm: \textgreater{}90\%                             \\
                                                           & 340-980 nm: \textgreater{}40\%                             \\
Readnoise (min)                                                  & 2.2 e$^{-}$                               \\ \hline
\multicolumn{2}{c}{Photometry}                                                                                          \\
FoV                                                        & 12$'$                                                        \\
Image scale (binning 1 $\times$ 1)            & 0$''$.45/pix                                                  \\
Limiting mag ($V$, 20 min, seeing $\sim$ 1$''$.1) & 22$^{\textrm{m}}$.5                                                       \\ \hline
\multicolumn{2}{c}{Stokes polarimetry}                                                                                  \\
FoV                                                        & 6$'$.4 $\times$ 6$'$.4                            \\
Image scale (binning 1 $\times$ 1)            & 0$''$.45/pix                                                  \\
Accuracy (14 mag, 20 min, seeing $\sim$ 1$''$)  & 0.6\%                                                      \\ \hline
\multicolumn{2}{c}{Long slit spectroscopy}                                                                              \\
Spectral range                                             & 400-720 nm                                                 \\
Spectral resolution                                        & $R$ $\sim$ 1000                                 \\
Slit size                                                  & 1$''$.7 $\times$ 12$'$                            \\
Monochromatic slit image (FWHM)                              & 3.5 pix                                                     \\
Reciprocal dispersion                                      & 0.2 nm/pix                                                  \\ \hline
\end{tabular}
\label{tab:magic}
\end{table}

\subsection{Baffles}
\label{sec:maths} % used for referring to this section from elsewhere

\begin{figure*}
    \centering
    \includegraphics[width=1\textwidth]{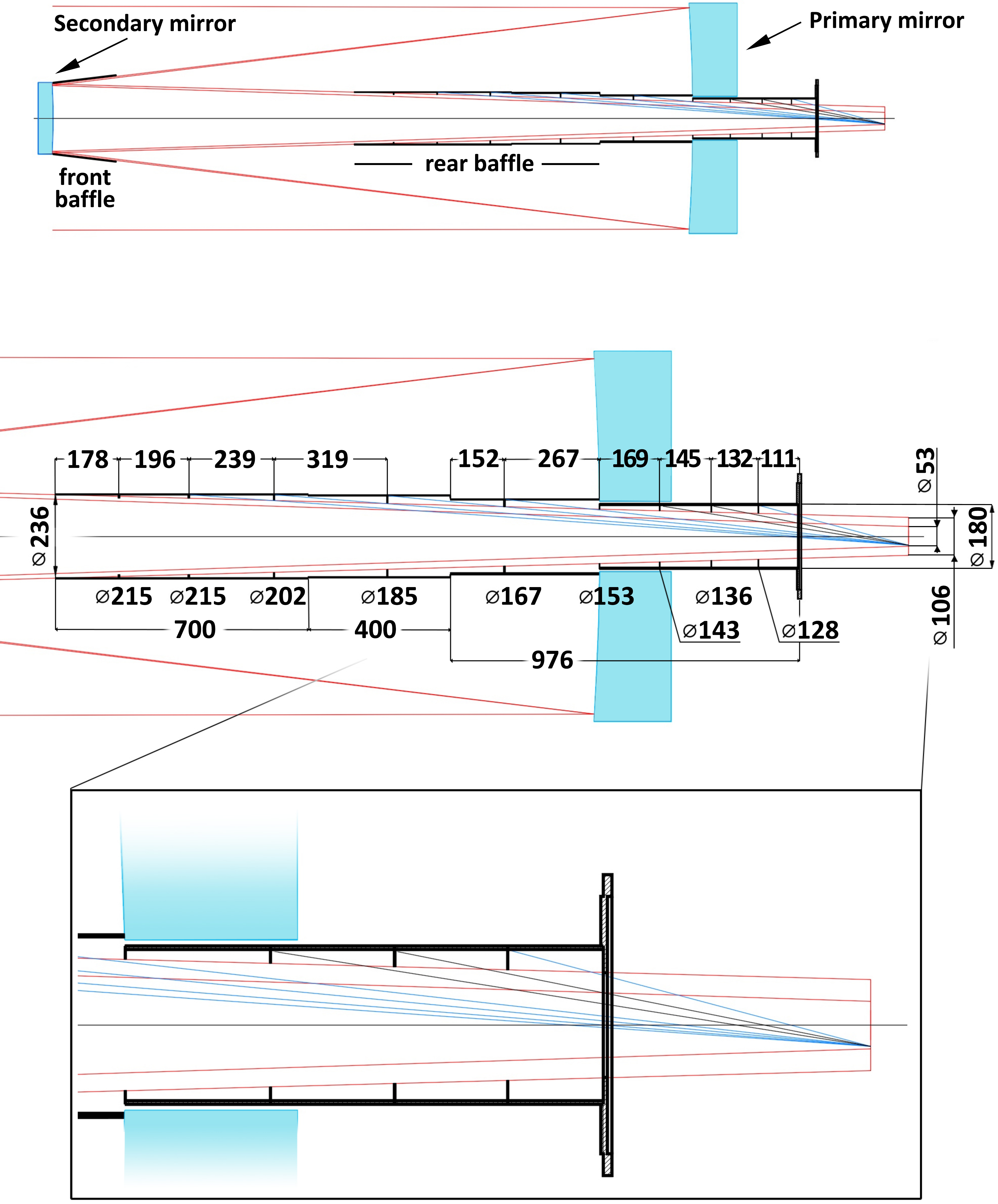}
    \caption{Optical scheme of the Zeiss-1000 telescope after modernization of the baffles. The top panel shows default front and rear baffles near secondary and primary mirrors with annular diaphragms installed in the rear one. Also, in the scheme (to the right to the rear baffle) there is an additional construction with diaphragms, which we installed through the main mirror of the telescope. The middle panel indicates the sizes of installed elements. The bottom panel shows the idea of arranging annular diaphragms described in \citep{idea}. Dimensions are in millimetres.}
    \label{fig:baffles_scheme}
\end{figure*}

The Zeiss-1000 telescope \citep{Z1000} is a Ritchey-Chr\'etien aplanat with two hyperbolic mirrors. Due to their design, Cassegrain telescopes are the most vulnerable to parasitic light incoming to the detector during observations. Baffles have been installed into the telescope as a system of two surfaces: a truncated cone and a cylinder (near the secondary and primary mirrors, respectively). They are shown on the top panel of Fig.~\ref{fig:baffles_scheme} and are called "front"{} and "rear" baffles. These are the default baffles originally installed into the telescope. This configuration provides an unvignetted field of $\diameter$106~mm ($\sim$27$'$) at the telescope focal plane. The baffles shield the detector from the most dangerous direct light and also prevent light reflected by the inner surface of the telescope tube from entering the field of view. However, the baffle near the main mirror causes additional scattered light, when direct light is reflected from the inner surface of the baffle during grazing incidence \citep{Maksutov}.

    \begin{figure*}
    \centering
    \includegraphics[width=1\textwidth]{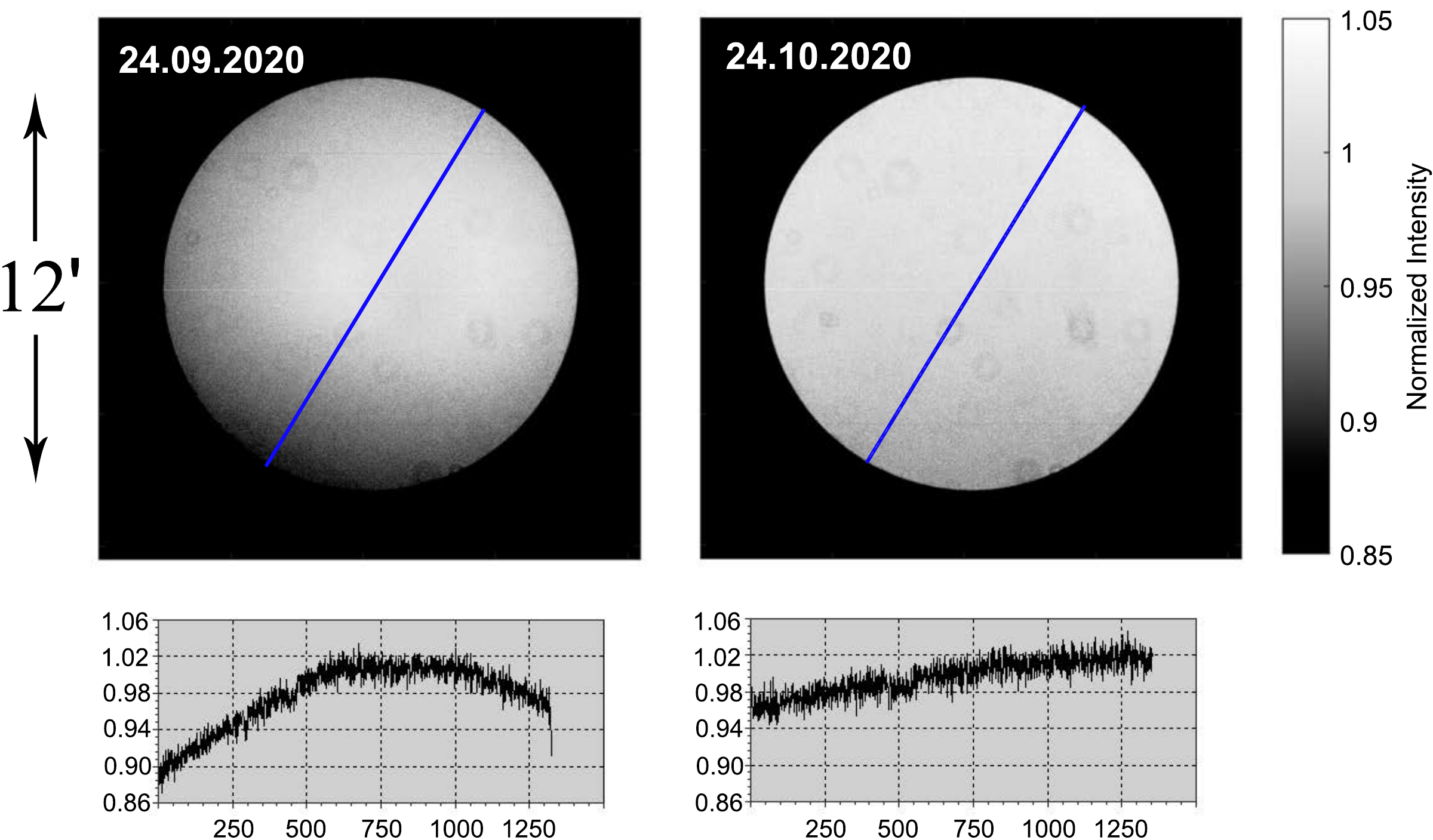}
    \caption{Comparison of the normalized flat field frames of the twilight sky {\textit{before}} (left) and {\textit{after}} (right) the installation of annular diaphragms in the default rear baffle and an additional tube, and blackening of the components. The cuts at the bottom correspond to the blue lines in the frames above. The horizontal axes of bottom cuts correspond to the pixel location along the $y$-axis of the frame (the length of the blue line in the angular measure corresponds to 12$'$). Frames obtained with a 250\AA-width SED700 filter.}
    \label{fig:baffles_flat}
\end{figure*}

Thus, due to light re-reflection on the inner surface of the rear baffle, a complex-shaped spot was formed on the detector, which, in its meaning, was an additive parasitic signal. We observed it as a drop of the intensity at the edges of the calibration flat-field frame of the order of $\sim$10\% (Fig.~\ref{fig:baffles_flat}, left panel). The maximum of this "bell"{} was shifting during observations depending on the position of the telescope tube, which introduced significant errors in obtaining flat field frames and data reduction. When processing scientific frames, scattered light cannot be taken into account, which worsens the accuracy of measurements of faint objects. Since we have a parasitic additive signal, the division of frames into the flat field also introduced a systematic error of about 10\% towards the edges of FoV and decreased the accuracy of high-precision photometric measurements. Moreover, the scattered light must contribute to the instrumental polarization, and its value is heterogeneous over the field.

Firstly, we performed the exact solution of the problem for calculating the optimum design of baffles \citep{baffles} for Zeiss-1000 to fully replace the original baffles. Yet, it appeared that the solution led to an unacceptably high linear obstruction coefficient (ratio of diameters of the widest baffle and the entrance pupil) $\eta \sim$ 0.46. Thus, we got out of the exact solution for a more acceptable design. 

%At first,\textbf{ we wanted} to make a new system of baffles (instead of the original ones). However, the exact solution of the problem for calculating the optimum design of baffles  for the Zeiss-1000 SAO RAS configuration led to an unacceptably high linear obstruction coefficient (ratio of diameters of the widest baffle and the entrance pupil) $\eta \sim$ 0.46.

To suppress unwanted light, we installed four annular diaphragms with an internal diameter of 185 to 215 mm inside the existing rear baffle (it consists of two parts, with a total height of 1100 mm) and painted the components with high absorption paint. We also made an additional cylindrical 976 mm high structure with five internal diaphragms, installed between the focal plane of the telescope and the default rear baffle, and passing through the central hole of the telescope's main mirror. A drawing of baffles with annular diaphragms installed inside in the Zeiss-1000 optical system of the SAO RAS is shown in Fig.~\ref{fig:baffles_scheme}. The idea of annular diaphragms for refractors was described earlier in \cite{idea} and is easily adapted to the design of a cylindrical baffle (the idea is visualized in the bottom panel of Fig.~\ref{fig:baffles_scheme}). Thus, diaphragms surround the useful light beam in the optical path and significantly reduce the level of unwanted light.

A comparison of flat field frames obtained from the twilight sky \textit{before} and \textit{after} blackening the baffle, installing painted diaphragms in it, and installing an additional structure with diaphragms is shown in Fig.~\ref{fig:baffles_flat} on the left and right panels, respectively. After the upgrade, the intensity of the flat field does not drop at the edges of the FoV, which indicates effective blocking of direct and scattered beams in the telescope tube.

%Refer back to them as e.g. equation~(\ref{eq:quadratic}).

\subsection{Rotator and offset guide}\label{subsec3}

The rotator, offset guide, calibration illumination as well as baffles are device-independent modules. Since the end of 2022, the rotator and offset guide are already being used in a test mode with the \MAGIC{} device and are available to be used with other devices installed at the Cassegrain focus. The calibration illumination is still under development. Below we briefly show their necessity and main features. The details of the rotator and offset guiding system will be described in the upcoming paper (Amirkhanyan et al. 2024, in prep.).

The Zeiss-1000 telescope was originally equipped with a manual rotator. We have upgraded the original Zeiss-1000 rotator by designing, manufacturing and assembling construction of large gear, worm reduction and a stepper motor with PCB control. Thus, this modification allows one to rotate remotely the devices installed in the Cassegrain focus to any given angle during the night, which makes observations using various methods much more efficient. The accuracy of the angle setting is $\sim$0$^{\circ}$.5.

The offset guide is designed to correct the position of the Zeiss-1000 telescope tube based on images from a guide digital camera mounted on a small gear platform into space inside the motorized rotator. An additional guiding module turned out to be necessary since the telescope's tracking error does not allow full-fledged exposures for several tens of minutes. Before the start of work on the production of the offset guide, the capabilities of the side telescope guide of the Zeiss-1000 telescope were tested. During guiding through the side telescope, we got a systematic drift of $\sim$2$''$.5 per hour, which became the prerequisite for the creation of the offset guide.

The rotation of the offset guide platform makes it possible to quickly find available stars for guiding in the %ring between two radii $R_1 \sim 10'.5$ and $R_2 \sim 15'.7$ relative to the centre of the telescope
FoV of the telescope at the Cassegrain focus. %The FoV is approximately $4' \times 3'$ and image scale --- about 0$''$.35/pix (these values depend from the secondary mirror position of the Zeiss-1000). 
The limiting magnitude of a star for guiding is $\sim$14 mag in $R$-band.

\begin{figure*}
    \centering
    \includegraphics[width=0.49\textwidth]{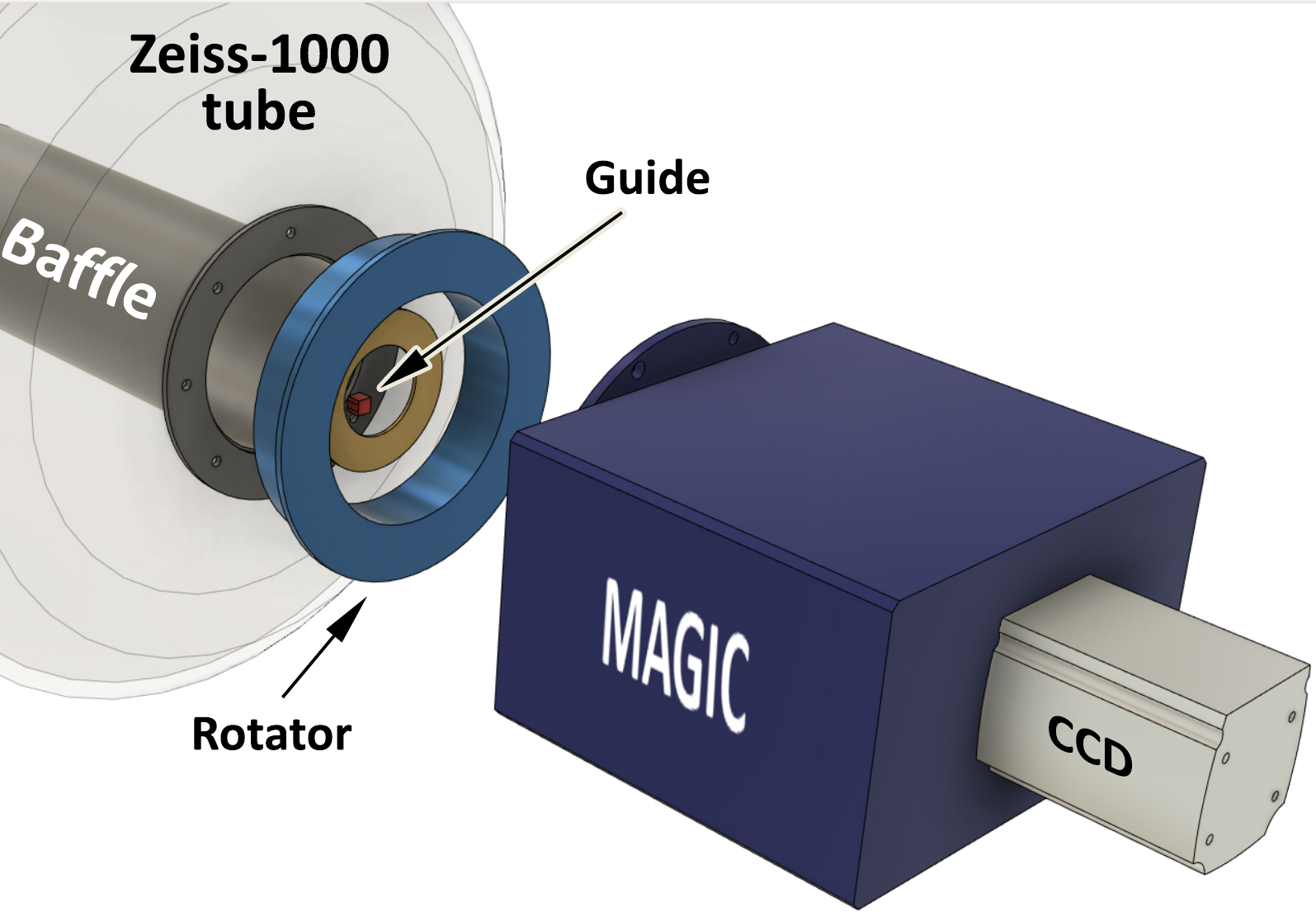}
    \includegraphics[width=0.45\textwidth]{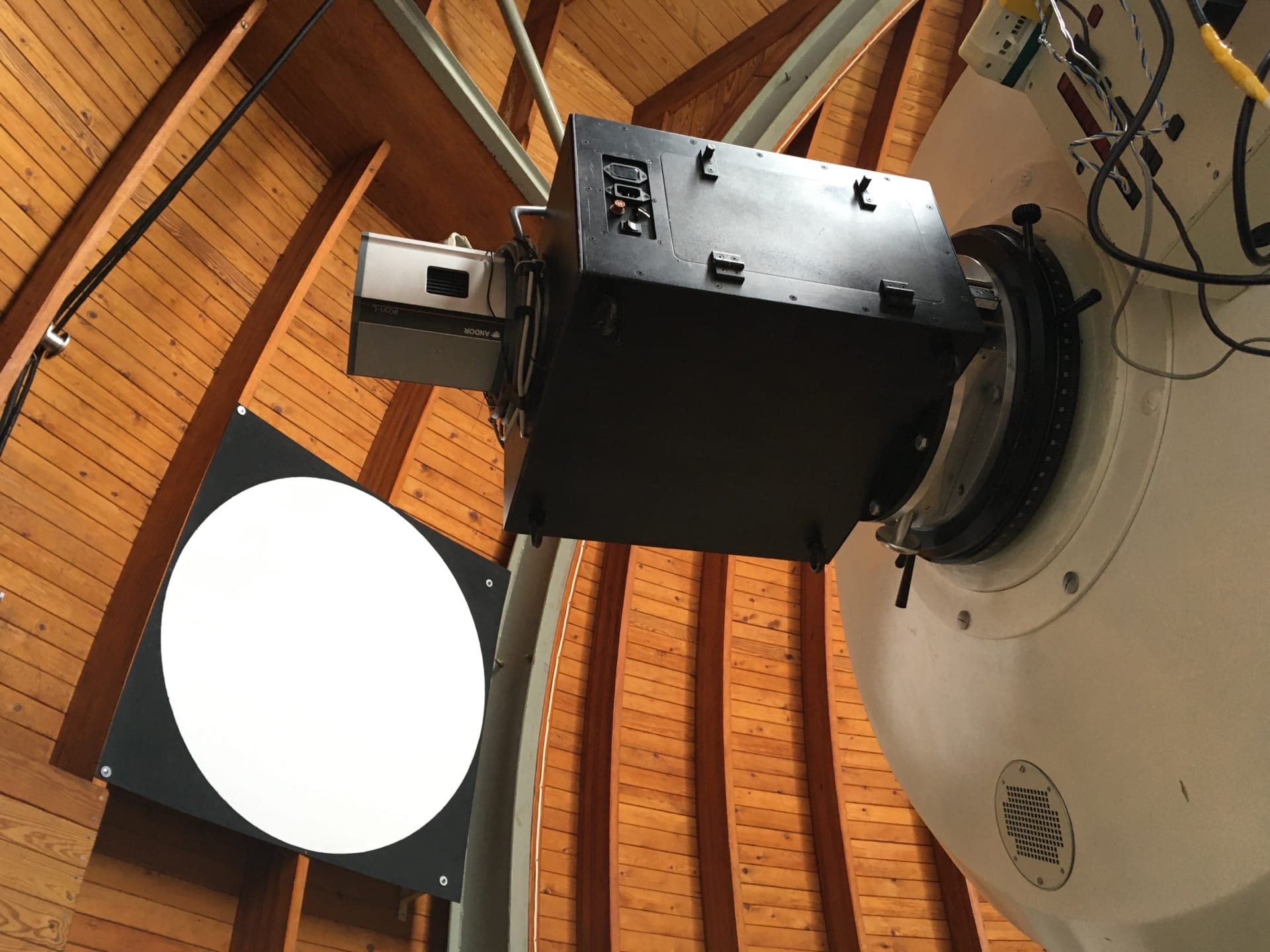}
    \caption{\MAGIC{} in the Cassegrain focus. \textit{Left}: An illustrative scheme with a transparent telescope tube. \textit{Right}: photo of \MAGIC{} and a round flat-field screen in the background.}
    \label{fig:MAGIC_flat}
\end{figure*}

\section{\MAGIC{} description}\label{sec3}

\begin{figure*}
    \centering
    \includegraphics[width=1.0\textwidth]{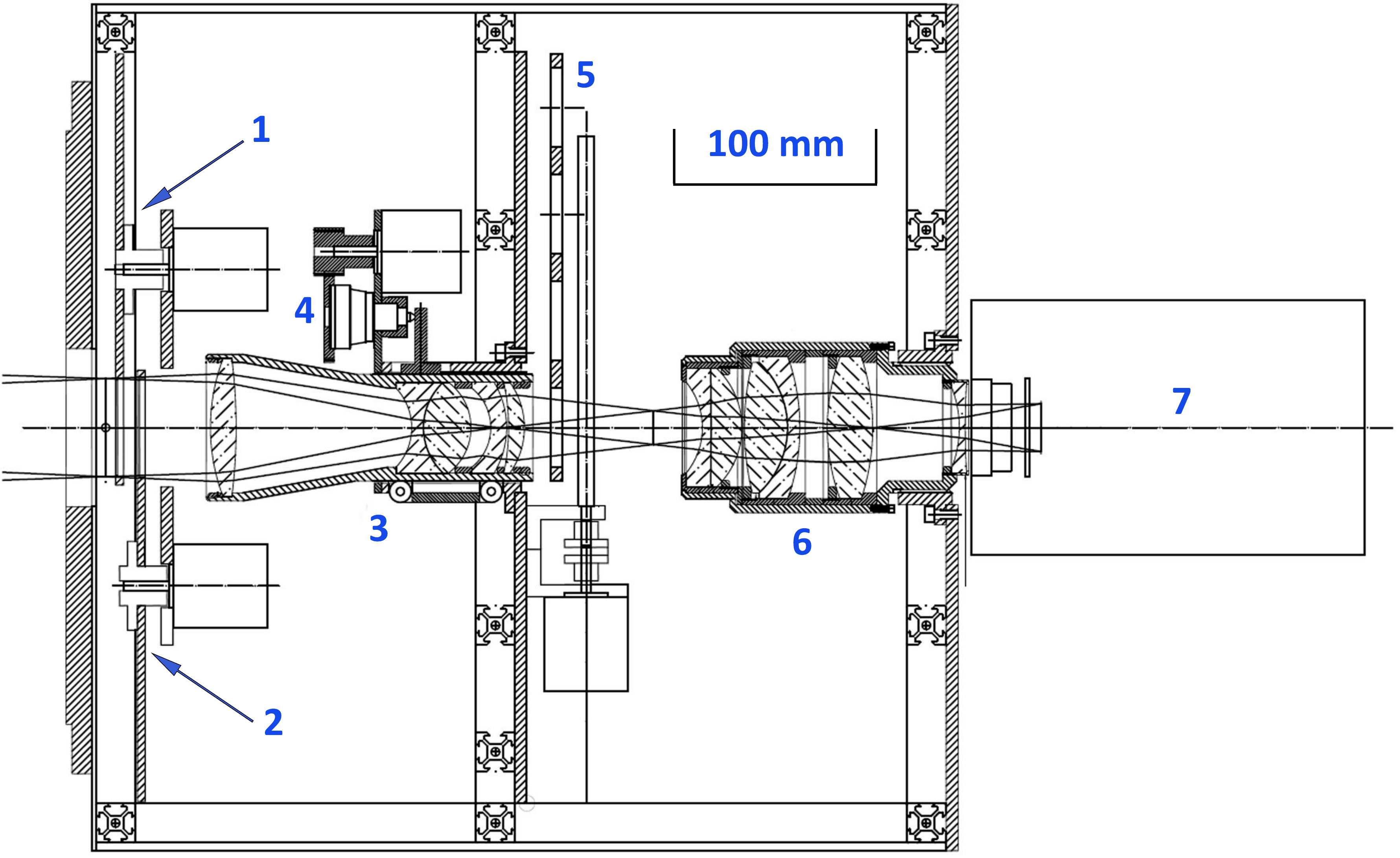}
     \caption{\MAGIC{} contents: 
     (1, 2) --- filter wheels;
     (3) --- collimator;
     (4) --- focusing mechanism of the collimator;
     (5) --- mode changing linear guide carriage;
     (6) --- camera; 
     (7) --- the CCD detector.}
    \label{fig:scheme}
\end{figure*}

The \texttt{MAGIC} device is a multi-mode focal reducer, 
%which makes it an extremely efficient instrument on a telescope, 
allowing a flexible response to changing weather conditions due to several observational modes: direct images, polarimetry and long-slit spectroscopy. 
%Indeed, when the weather worsens, an ordinary polarimeter does not allow spectral observations to be made, and its efficiency comes to naught. 
%At the same time, several observation modes are implemented in \texttt{MAGIC}: direct images, polarimetry and long-slit spectroscopy -- which technically allows solving various observational tasks in accordance with the current weather conditions.
\texttt{MAGIC} is installed in the Cassegrain focus of the 1-m Zeiss-1000 telescope and works in conjunction with the components of the optical system described earlier (see Fig. \ref{fig:MAGIC_flat}), but does not depend on them. The weight of the device without a CCD detector is 23 kg, and the size is 430$\times$440$\times$265 mm.

The device is designed for an input focal ratio of $F/12.5$ and, due to the collimator and camera, reduces it to $F/6.1$, which solves the problem of oversampling for typical modern CCDs in the focus of Cassegrain telescopes and provides an advantage for observing faint extended objects. %On the other hand, since the collimator forms a parallel beam, there is an optimal possibility of using dispersive elements in it, as well as polarization analyzers.

\begin{figure}
    \centering
    \includegraphics[width=0.8\columnwidth,angle=90]{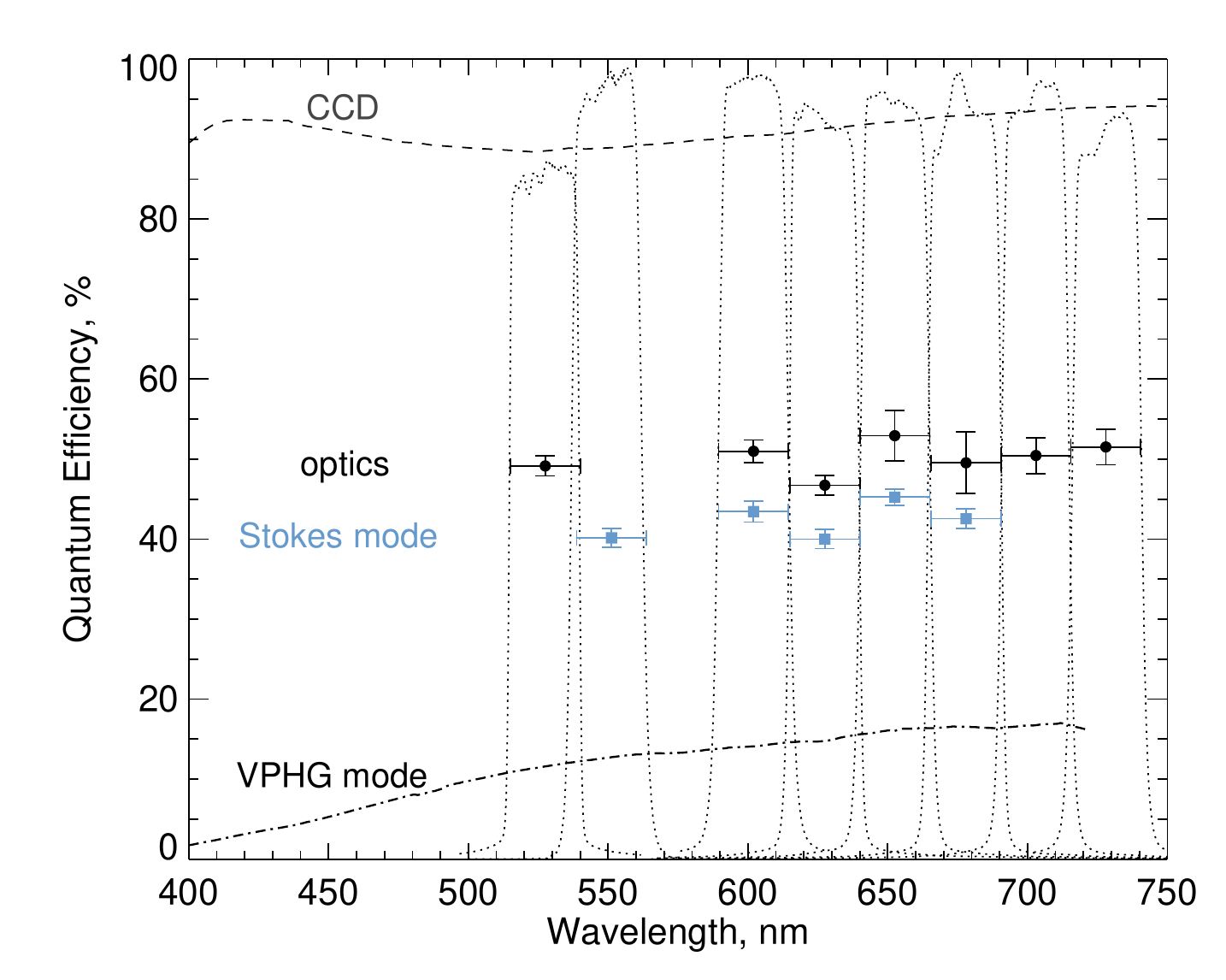}
    \caption{QE of the system \MAGIC{}+telescope+CCD. Filled black circles with error bars mark transmission measurements of the \MAGIC{} with the Zeiss-1000 telescope mirrors and CCD. Blue squares present the same including the transmission of the quadruple Wollaston prism. The dash-dotted line presents the QE in the spectral mode with the VPHG (including optics+telescope+CCD). The dashed line also shows the quantum efficiency of the CCD for this spectral range. The pass-bands of the medium-band SED filters used to measure QE are plotted with a dotted line.}
    \label{fig:qe}
\end{figure}

\begin{figure*}
    \centering
    \includegraphics[height=0.60\textwidth]{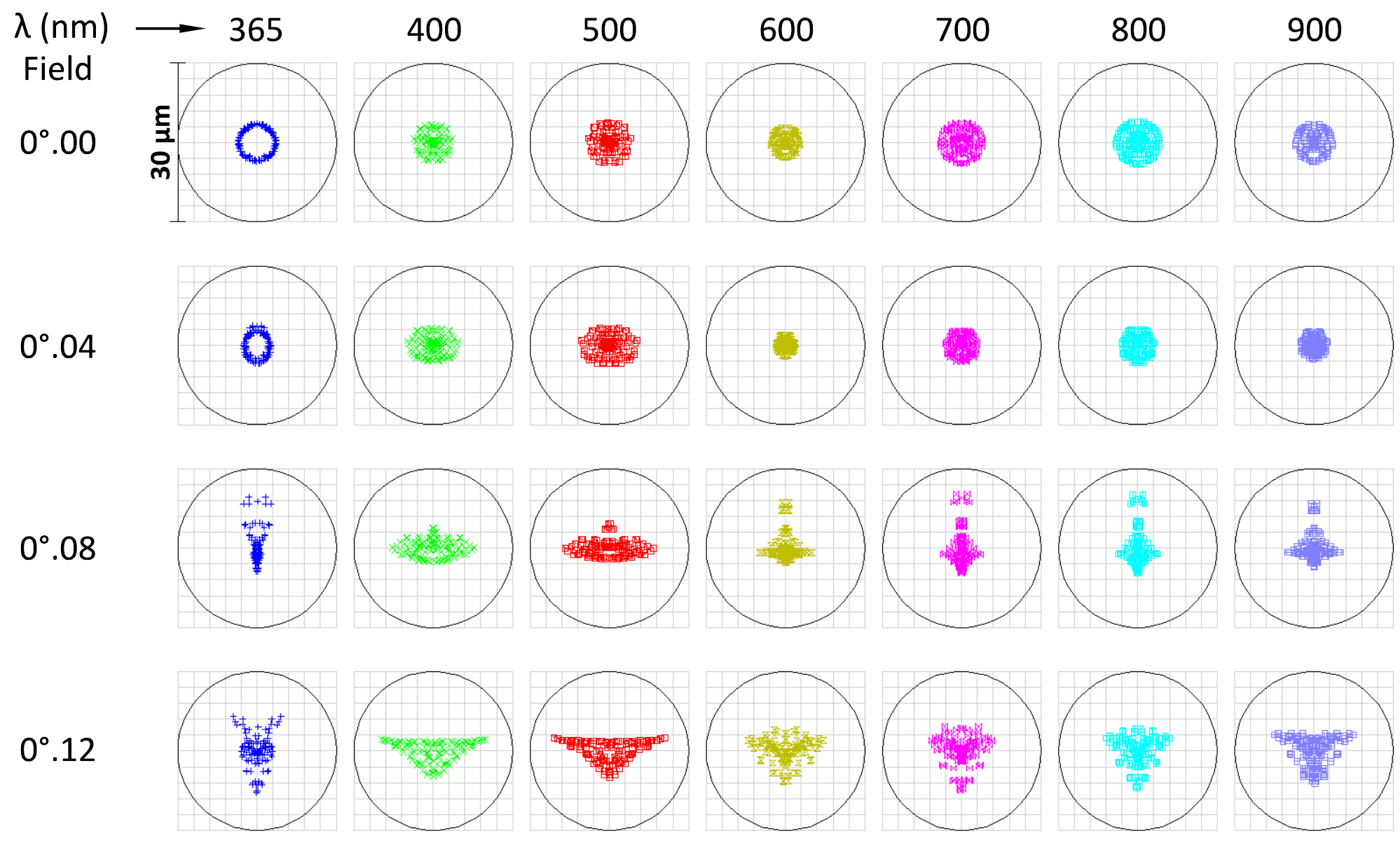}
    \caption{Spot Diagram. Circle diameter --- 30 microns = 1$''$.}
    \label{fig:SPOT}
\end{figure*}

\subsection{Optical design}\label{subsec5}

The optical part of the \texttt{MAGIC} focal reducer consists of a field lens, a collimator and a camera lens. The scheme is shown in Fig. \ref{fig:scheme}. The collimator is a 5-lens apochromat with a focal length of 220~mm and forms the exit pupil of the system. The camera lens is a 6-lens apochromat with a focal length of 109~mm, which focuses the resulting image on the CCD detector. All optical surfaces have an anti-reflective coating, which ensures transmission of each lens $>$80\%. The integral transmission\footnote{The quantum efficiency of \MAGIC{} optics and observational modes was measured by on-sky standard stars in medium-band filters with the known pass-bands%, taking into account the obstruction of the primary mirror by baffles and measured gain characteristic of CCD
.} of the focal reducer optics considering the reflection coefficient of telescope mirrors and CCD efficiency %\textcolor{blue}{\textbf{was estimated by on-sky standard stars observations in medium-band filters with the known pass-bands and}}
is shown in Fig. \ref{fig:qe} and is %\textcolor{blue}{\textbf{The average quantum efficiency of the device is}}
QE $\sim$ 50\%. %\textcolor{blue}{\textbf{We also plotted the QE in the polarimetric and spectroscopic modes in Fig. \ref{fig:qe} (see details in Sec. \ref{subsec10} and \ref{subsec11}).}}

The optomechanics of the device allow introducing the movable optical elements into the optical path. The optical filters can be additionally set in front of the collimator. Also, between the collimator and the camera, a volume phase holographic grism (VPHG) and a double Wollaston prism can be introduced into the parallel beam by moving the linear guide carriage perpendicular to the central axis of the device; it is also allowed to install other optical elements on the carriage.

\begin{figure}
    \centering
\includegraphics[scale=0.35, angle=90]{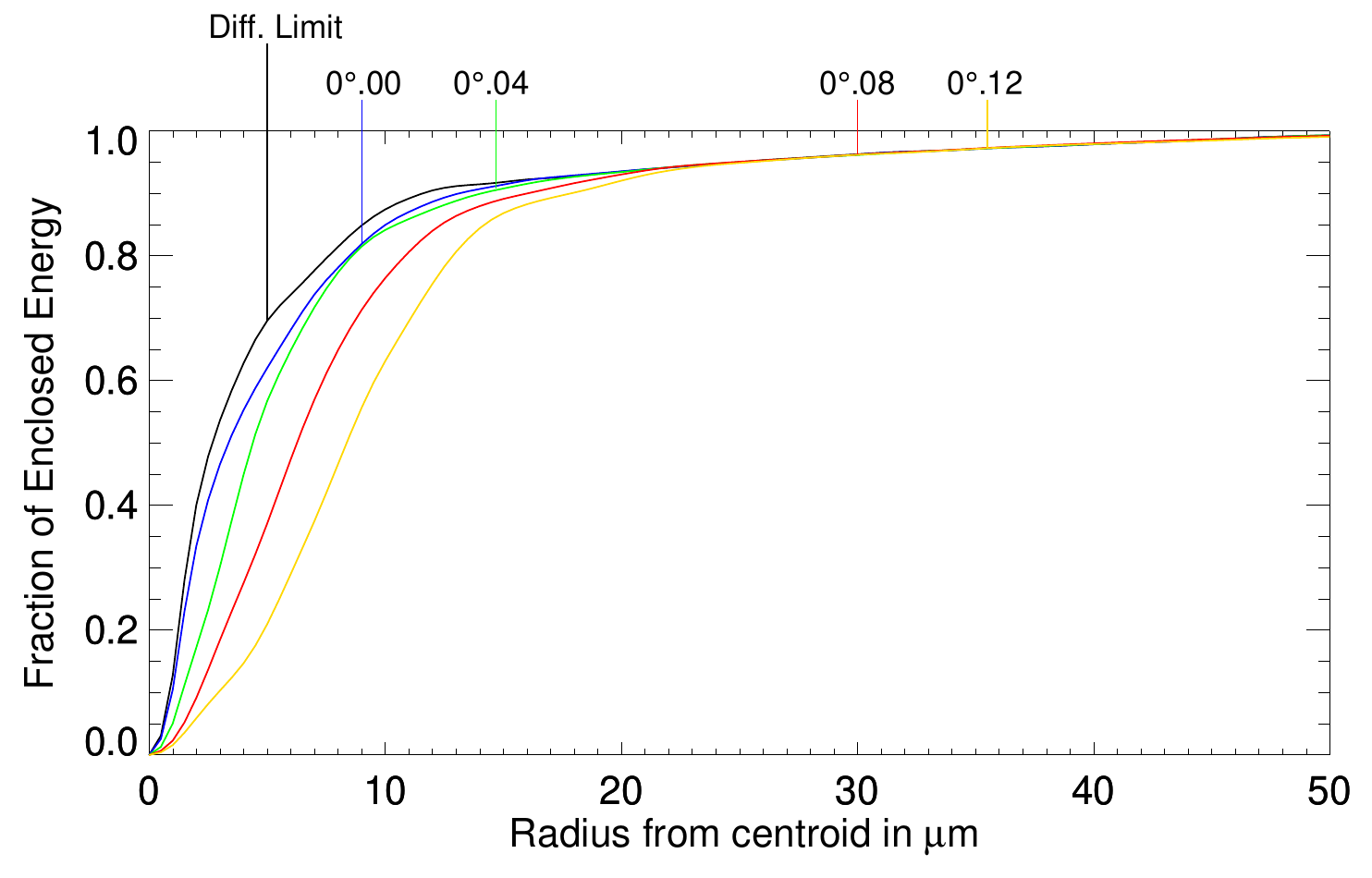}
    \caption{Polychromatic Encircled Energy.}
    \label{fig:PSF}
\end{figure}

The optical design of \MAGIC{} was calculated in the \texttt{ZEMAX} software environment.
Spot diagram in Fig. \ref{fig:SPOT} shows how the calculated image of a point source looks like for a series of wavelengths from 365~nm to 900~nm at various distances from the central axis of the device from 0$^{\circ}$ to $0^{\circ}.12$.
The calculated polychromatic encircled energy (the fraction of the total energy in the point spread function) is shown in Fig. \ref{fig:PSF}. The quality of the image formed by the optics is no worse than 10 $\mu$m in the plane of the CCD detector, which corresponds to FWHM $\sim$ 0$''$.3.

%\subsubsection{This is an example for third level head---subsubsection head}\label{subsubsec2}

\subsection{Electro-mechanical scheme}\label{subsec6}

In the \MAGIC{} scheme (Fig.~\ref{fig:scheme}), the light from the telescope passes through the filter wheels (1) and (2). Each wheel has 9 positions for installing filters with a diameter of no more than 50~mm and a thickness of no more than 5~mm. The first wheel, in addition to optical filters, also includes:

\begin{itemize}
    \item \textit{slit} --- long slit (width $1''.7$, linear width --- 0.11~mm)
    \item \textit{mask} --- mask for the Wollaston prism (angular dimensions --- $6'.4\times 6'.4$, linear dimensions --- 25$\times$25~mm)  
    \item \textit{dots} --- a matrix of 8$\times$8 pinholes with a diameter of 0.1~mm and a step of 3~mm for focusing optics and estimating geometric distortions in polarimetry mode (linear dimensions --- 25$\times$25~mm)
\end{itemize}
Zero position in each wheel is always empty, and given the constant presence of $slit$, $mask$ and $dots$, we have 13 positions to install the necessary replaceable filters.

Next, there is the collimator (3) with the focusing mechanism (4).
In the heart of \MAGIC{} is the mode-changing linear guide carriage for 4 positions (5) with the VPH-grism and the Wollaston prism.
The switching time between the adjacent carriage positions is 1 min.
After the mode carriage light comes through the camera (6) to the CCD detector (7).

To change the configuration, \texttt{MAGIC} has 4 stepper motors: two --- for rotating the filter wheels (1) and (2) and two more --- for the collimator focusing mechanism (4) and moving the linear guide carriage (5). 
%The stepper motors are controlled via an AVR~ATmega8535 microprocessor-based control board. A compact industrial computer MR3253S-00F with the Windows 7 operating system is installed inside the \texttt{MAGIC} case.
The control program from the onboard PC sends commands to the ATmega8535 microprocessor, which controls the configuration and activates the mechanics of the device.
The motors are controlled via the serial port from the graphical user interface (Fig.~\ref{fig:remote_control}).

\subsection{CCD characteristics}\label{subsec7}

Andor iKon-L 936 CCD system with a BEX2-DD type 2048~$\times$~2048~pix E2V CCD42-40 with a pixel size of 13.5~$\times$~13.5~$\mu$m is used as a detector. The mass of the CCD system is 7 kg. The quantum efficiency of this device is $>$90\% in the range of 400-850 nm (see Fig.~\ref{fig:qe}) and not less than 40\% in the range of 340-990 nm, which is the working spectral range of \texttt{MAGIC} due to its optics.
We use default air cooling, which makes it possible to conduct observations with a CCD temperature of about --80$^\circ$C.

The laboratory measurements of the gain value for the 1$\times$1 binning mode used in the observations are presented in Table \ref{tab:gain}. We use two gain modes 'low'\ ($\times1$) and 'high'{} ($\times4)$, as well as three readout rates for full frame -- 'fast'\ (4~sec), 'norm'\ (9~sec) and 'slow'\ (90~sec). The value of the measured readnoise for these modes is shown in Table \ref{tab:noise}. Note here that the measured values of CCD gain and readout noise differ significantly from the values provided by the manufacturer (19-28\% less than the declared gain and 26-45\% less than the declared readnoise, depending on the mode).

% \begin{table}[]
% \begin{tabular}{|cc|ccc|}
% \hline
% \multicolumn{2}{|c|}{{ }}                                                                                                     & \multicolumn{3}{c|}{{ rate}}                                                                                                                   \\ \cline{3-5} 
% \multicolumn{2}{|c|}{\multirow{-3}{*}{\begin{tabular}[c]{@{}c@{}}BIN 1$\times$1 (2048 $\times$ 2048 pix) \\ scale 0$''$.45/pix\end{tabular}}} & \multicolumn{1}{c|}{{fast (3.0 MHz)}} & \multicolumn{1}{c|}{{norm (1.0 MHz)}} & {slow (0.1 MHz)} \\ \hline
% \multicolumn{1}{|c|}{}                                                                   & high ($\times4$)                                             & \multicolumn{1}{c|}{0.89}                                  & \multicolumn{1}{c|}{0.84}                                       &0.84                                        \\ \cline{2-5} 
% \multicolumn{1}{|c|}{\multirow{-2}{*}{GAIN}}                                             & low ($\times1$)                                              & \multicolumn{1}{c|}{3.0}                                  & \multicolumn{1}{c|}{2.8}                                  & 2.8                                  \\ \hline
% \end{tabular}
% \caption{Measuring the value of the gain for various modes CCD Andor iKon-L 936}
% \label{tab:gain}
% \end{table}

\begin{table*}
\caption{Measurement of the gain value for various modes Andor iKon-L 936 CCD }
\centering
\begin{tabular}{ccccc}
\hline
\multicolumn{2}{c}{\multirow{2}{*}{\begin{tabular}[c]{@{}c@{}}Binning 1$\times$1 \\ 2048 $\times$ 2048 pix \end{tabular}}} & \multicolumn{3}{c}{rate}                         \\ \cline{3-5} 
\multicolumn{2}{c}{}                                                                                                                   & fast (3.0 MHz) & norm (1.0 MHz) & slow (0.1 MHz) \\ \hline
\multirow{2}{*}{GAIN}                                                & high ($\times$4)                                                & 0.89           & 0.84           & 0.84           \\
                                                                     & low ($\times$1)                                                 & 3.0            & 2.8            & 2.8            \\ \hline
\end{tabular}
%\label{tab:gain}
\label{tab:gain}
\end{table*}

% \begin{table}[]
% \begin{tabular}{|cc|ccc|}
% \hline
% \multicolumn{2}{|c|}{{ }}                                                                                                     & \multicolumn{3}{c|}{{ rate}}                                                                                                                   \\ \cline{3-5} 
% \multicolumn{2}{|c|}{\multirow{-2}{*}{\begin{tabular}[c]{@{}c@{}}Readnoise of CCD (in e$^{-}$) \\ with temperature $\sim$ -- 80 degrees\end{tabular}}} & \multicolumn{1}{c|}{{fast (3.0 MHz)}} & \multicolumn{1}{c|}{{norm (1.0 MHz)}} & {slow (0.1 MHz)} \\ \hline
% \multicolumn{1}{|c|}{}                                                                   & high ($\times4$)                                             & \multicolumn{1}{c|}{6.7 $\pm$ 0.03}                                  & \multicolumn{1}{c|}{4.8 $\pm$ 0.01}                                       &2.2 $\pm$ 0.01                                        \\ \cline{2-5} 
% \multicolumn{1}{|c|}{\multirow{-2}{*}{GAIN}}                                             & low ($\times1$)                                              & \multicolumn{1}{c|}{11.3 $\pm$ 0.11}                                  & \multicolumn{1}{c|}{5.9 $\pm$ 0.06}                                  & 2.7 $\pm$ 0.07                                  \\ \hline
% \end{tabular}
% \caption{Measuring of read noise for various Andor iKon-L 936 CCD modes}
% \label{tab:noise}
% \end{table}

\begin{table*}
\caption{Measurement of the readnoise for various Andor iKon-L 936 CCD modes}
\centering
\begin{tabular}{ccccc}
\hline
\multicolumn{2}{c}{{ }}                                                                                                     & \multicolumn{3}{c}{{ rate}}                                                                                                                   \\ \cline{3-5} 
\multicolumn{2}{c}{\multirow{-2}{*}{\begin{tabular}[c]{@{}c@{}}CCD readnoise (in e$^{-}$) \\ under --80$^\circ$C\end{tabular}}} & \multicolumn{1}{c}{{fast (3.0 MHz)}} & \multicolumn{1}{c}{{norm (1.0 MHz)}} & {slow (0.1 MHz)} \\ \hline
\multicolumn{1}{c}{}                                                                   & high ($\times4$)                                             & \multicolumn{1}{c}{6.7 $\pm$ 0.03}                                  & \multicolumn{1}{c}{4.8 $\pm$ 0.01}                                       &2.2 $\pm$ 0.01                                        \\  
\multicolumn{1}{c}{\multirow{-2}{*}{GAIN}}                                             & low ($\times1$)                                              & \multicolumn{1}{c}{11.3 $\pm$ 0.11}                                  & \multicolumn{1}{c}{5.9 $\pm$ 0.06}                                  & 2.7 $\pm$ 0.07                                  \\ \hline 
\end{tabular}
\label{tab:noise}
\end{table*}

\begin{figure*}
    \centering
    \includegraphics[width=0.33\textwidth, angle=90]{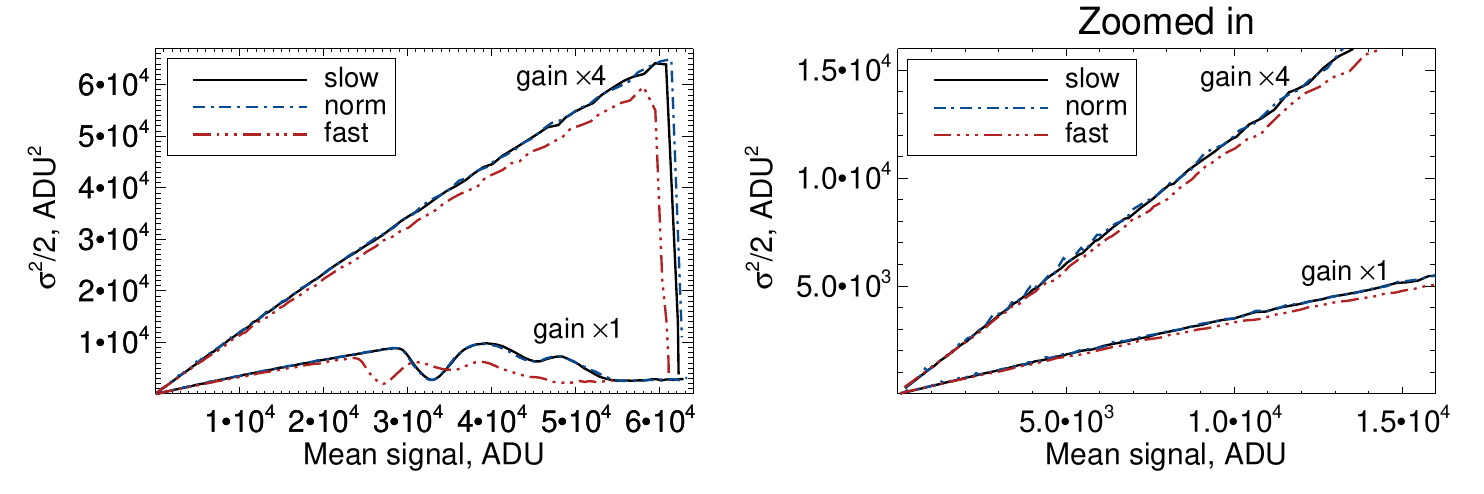}
    \caption{The gain factor is determined from the slope of the dependence of half of the variance of counts on the average value of signal accumulation. \textit{Left}: dependencies are presented for all gain modes and readout rates used in the observations. \textit{Right}: zoomed in on the same dependencies.}
    \label{fig:gain}
\end{figure*}

% \begin{table}[]
% \centering
% \caption{Measurements of the values of gain and readout noise for various modes CCD Andor iKon-L 936 in 1$\times$1 (2048$\times$2048 pix, 0$''$.45/pix scale) binning under $-80$ degrees temperature. Readnoise of CCD is given in e$^{-}$. }
% \begin{tabular}{cccc}
% \hline
%                                 & \multicolumn{1}{c}{} & \multicolumn{1}{c}{Gain} & Readout noise   \\ \hline
% \multirow{2}{*}{fast (3.0 MHz)} & $\times$1            & 3.0                      & 11.3 $\pm$ 0.11 \\ \cline{2-4} 
%                                 & $\times$4            & 0.89                     & 6.7 $\pm$ 0.03  \\ \hline
% \multirow{2}{*}{norm (1.0 MHz)} & $\times$1            & 2.8                      & 5.9 $\pm$ 0.06  \\ \cline{2-4} 
%                                 & $\times$4            & 0.84                     & 4.8 $\pm$ 0.01  \\ \hline
% \multirow{2}{*}{slow (0.1 MHz)} & $\times$1            & 2.8                      & 2.7 $\pm$ 0.07  \\ \cline{2-4} 
%                                 & $\times$4            & 0.84                     & 2.2 $\pm$ 0.01  \\ \hline
% \end{tabular}
% \end{table}

It is significant that there is a misconception that the statistics of counts (analogue digital units, ADU) in CCDs correspond to Poisson ones. This assumption is laid down when determining the gain factor of the analogue-to-digital converter of the CCD registration path \citep{Howell}. However, as can be seen in Fig.~\ref{fig:gain} (and especially on the right panel, where the range of the graph is zoomed in), the dependence of the counts variance on the average registered signal is different from a strictly linear law. There are periodic fluctuations around a linear dependence.%, the nature of which remains unknown to us.
~We assume that this is a feature of thick silicon CCD detectors with deep depletion technology.

Also, based on the measurements in Fig.~\ref{fig:gain}, we can identify the working ranges of ADU accumulation for observations in various modes (for gain $\times1$\ and $\times4$) of CCD iKon-L 936, where the signal dispersion behaves in the most acceptable way. It can be concluded that for ($\times$1) low gain mode it is not worth accumulating a signal of more than $\sim$20k ADU.

\begin{figure*}
    \centering
    \includegraphics[width=0.65\textwidth, angle=90]{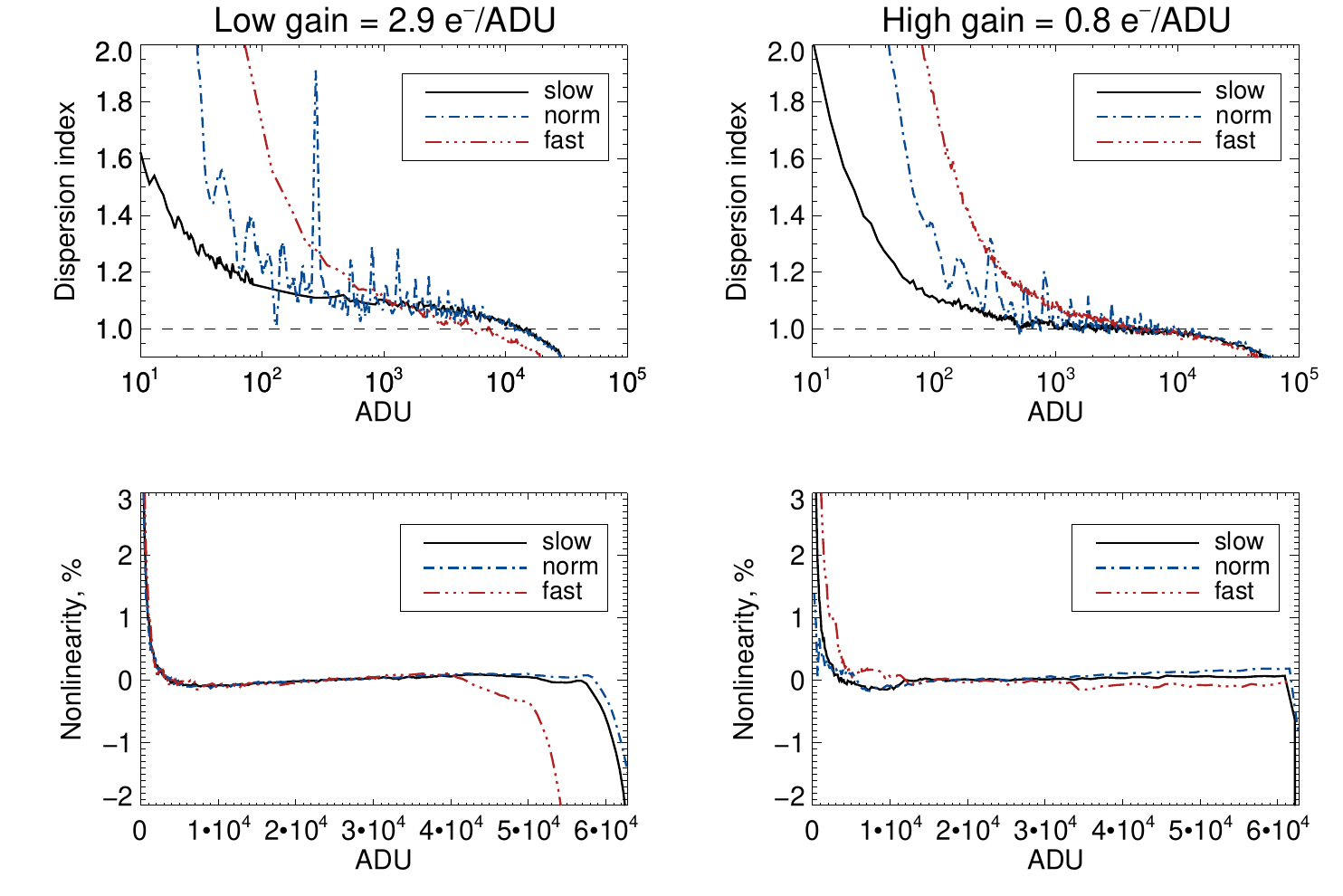}
    \caption{Measurement of CCD characteristics for all gain modes (\textit{left}: $\times$4 high, \textit{right}: $\times$1 low) and readout rates. The top panel shows the dependence of the dispersion index on signal accumulation. The lower panel shows the level of non-linearity of signal registration in the entire range of accumulations.}
    \label{fig:FANO}
\end{figure*}

On the other hand, for astronomical observations, the particular interest is the registration of weak signals, whose statistics are distorted by the readout noise introduced by the electronics. To study the distortion of counts statistics, a test criterion is used using the dispersion index, the so-called Fano factor~\citep{Fano1947}. The application of the method to CCD studies is described in detail by~\cite{Afanasieva2016}. By definition, the dispersion index is the ratio of the variance of counts to the average value of the registered signal. For a Poisson distribution, this ratio is equal to one, and this corresponds only to a certain range of registered values. Fig.~\ref{fig:FANO} shows graphs of the dependence of the dispersion index on the magnitude of the registered signal in different modes for the iKon-L 936 CCD. The left and right panels correspond to two gain modes -- ($\times1$) low and ($\times4$) high respectively. These studies also provide insight into the optimal choice of exposure time in order to minimize the distortion of counts statistics when observing astrophysical objects using the \texttt{MAGIC} focal reducer. According to the measurements, the best fit to the Poisson statistics is achieved when the signal is accumulated in the ($\times$1) low gain mode at a 'slow' readout rate from about a few hundred to $\sim$10k ADU.

Note here that for both CCD gain modes used ($\times$1 and $\times$4) for the 'norm' readout rate, 'sawtooth' beats of the dispersion index are observed. We keep in mind this negative feature during observations.

Also in Fig.~\ref{fig:FANO} on the bottom panels there are measurements of the deviation from signal linearity, which do not exceed 0.5\% in the entire range of signal accumulations used in observations.

CCDs with a thick, deep-depletion silicon substrate provide high spectral sensitivity of the detector even in the 1 $\mu$m region. A powerful advantage of the iKon-L 936 CCD is the complete absence of interference noise in the red part of the spectrum. Under laboratory conditions, we exposed the CCD illuminated with various wavelengths and could not detect the contribution of the interference pattern, so-called fringes. Thus, this CCD allows one to efficiently provide research in the red part of the spectrum at high sensitivity. Additional information about the peculiarities of CCD images in the near-infrared band is given in Appendix \ref{secA1}.

% We noticed inhomogeneities of the order of 2\% only when exposed to illumination with $\lambda$\ = 1050 nm, while the noise did not correspond to the classical interference pattern. An example of a frame normalized to average intensity is shown in Fig.~\ref{fig:micron}. These were rounded inhomogeneities, the nature of which can be different (from manifestations of optic at $\lambda > 1$\ $\mu$m to thermal noise from electrodes). Thus, this CCD allows useful research in the red part of the spectrum, without complicating observations with various interferences like fringes, at high sensitivity.

% \begin{figure}
%     \centering
%     \includegraphics[width=1.0\textwidth, angle=0]{pics/ccd_fig/flat1050.eps}
%     \caption{A frame (2048 $\times$ 2048 pix) normalized to average intensity of exposure to illumination at a wavelength of 1050 nanometers.}
%     \label{fig:micron}
% \end{figure}

\subsection{Remote control}\label{subsec8}

\begin{figure*}
    \centering
    \includegraphics[width=1.0\textwidth]{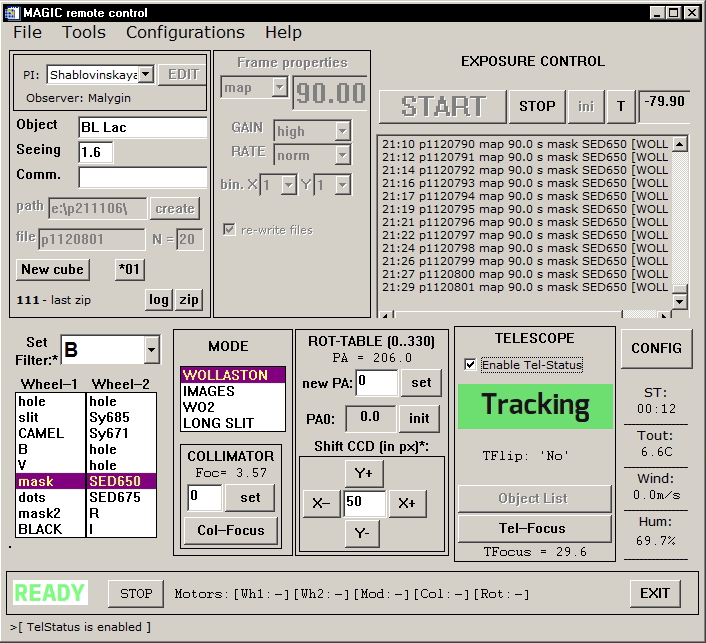}
    \caption{\MAGIC{} control interface.}
    \label{fig:remote_control}
\end{figure*}

\begin{figure*}
    \centering
    \includegraphics[width=1.0\textwidth]{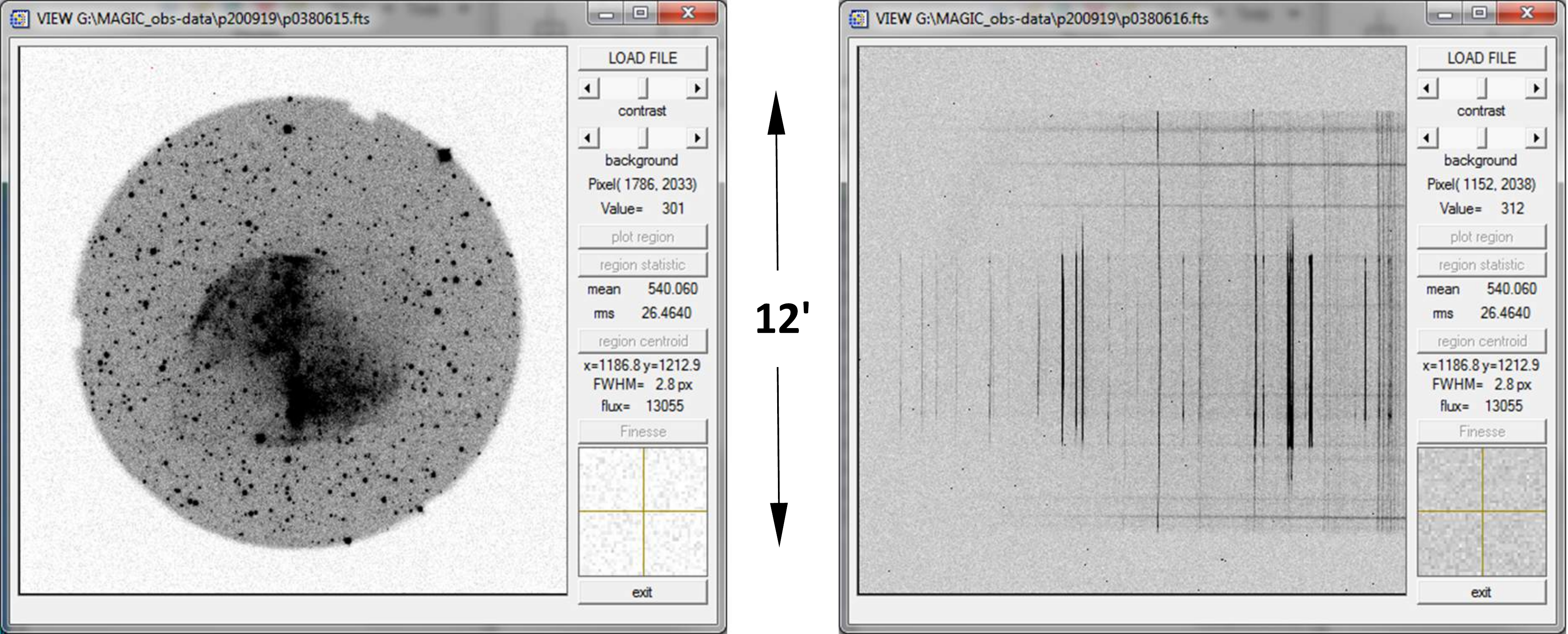}
    \caption{Viewer interface with frames of the M27 planetary nebula in photometric (\textit{left}, $t_{\rm exp}$ = 10~s in $R_{\textrm{C}}$-band) and spectral (\textit{right}, $t_{\rm exp}$ = 600~s) modes. Direct image FoV is 12$'$ $\times$ 12$'$, slit height is 12$'$, slit width is $1''.7$, the wavelength range is 340-740 nm. The frames colours are inverted.}
    \label{fig:viewer}
\end{figure*}

The control of the device, including the rotator, guide and CCD, is implemented through several compact computers installed on the telescope, which allows remote observations.
In observations, we use network access to the onboard computer MR3253S-00F (with Windows 7 as the operating system) made by LEX COMPUTECH in the remote desktop format. The control interface is a graphical shell in the IDL environment \texttt{MAGIC remote control}, a screenshot of which is shown in Fig.~\ref{fig:remote_control}. The upper half of the interface is used to control the CCD detector and edit the information recorded in the FITS header during the observations; the lower half is used to control the \texttt{MAGIC} (setting the observation mode, focusing the collimator, and orientation) and some telescope functions (small tube shifts and focusing). At the end of each exposure, the resulting FITS file is opened for analysis in the FITS-viewer (see Fig.~\ref{fig:viewer}) --- here the observer traditionally controls the levels of accumulation and the quality of each frame. Note here that the image in the viewer is flipped along the RA axis. 

%A similar compact computer permanently installed on the Zeiss-1000 telescope serves to control the guiding and rotator.

\section{Observation Modes}\label{sec4}

\subsection{Photometry}\label{subsec9}

The photometric mode of observations with the \texttt{MAGIC} device makes it possible to obtain direct images using various light filters, which are introduced into the beam by means of two wheels. The size of the FoV is limited by the size of the round filter and is $\sim$12$'$. Note that for photometry, as well as in other observation modes, we use 1$\times$1 CCD binning, which gives an image scale of 0$''$.45/pix and satisfies the Kotelnikov-Nyquist theorem (sampling allows us to accurately restore the PSF-profile). The device uses narrow-band and medium-band interference SED filters\footnote{Manufactured by Edmund Optics, \url{https://www.edmundoptics.com/}. Included in SCORPIO-2 filter set \url{https://www.sao.ru/hq/lsfvo/devices/scorpio-2/filters\_eng.html}.} (the bandwidths of the SED filters used to measure QE are shown in Fig.~\ref{fig:qe}), as well as broadband glass filters %\textit{BVR${_C}$I}$_{C}$ 
\textit{BVR}$_{\textrm{C}}$\textit{I}$_{\textrm{C}}$ of the Johnson-Cousins system \citep{1990Bessell}. In the case of the broadband filters, for converting instrumental quantities %(lowercase letters) 
into \textit{standard photometric system} %(capital letters) 
the following equations were constructed neglecting the second-order extinction coefficients:

\begin{equation}
\label{photosystem}
 \begin{array}{c}
  B = b + 0.12^{\pm 0.022}(B - V) + 22.43^{\pm 0.014}\\
  V = v - 0.23^{\pm 0.023}(B - V) + 22.78^{\pm 0.015}\\
  R_{\textrm{C}} = r + 0.22^{\pm 0.043}(V - R_{\textrm{C}}) + 22.75^{\pm 0.017}\\
  I_{\textrm{C}} = i + 0.05^{\pm 0.022}(V - I_{\textrm{C}}) + 22.23^{\pm 0.019}
 \end{array}
\end{equation}

where \textit{B}, \textit{V}, \textit{R}$_{\textrm{C}}$, \textit{I}$_{\textrm{C}}$ are standard magnitudes in \textit{B}-, \textit{V}-, \textit{R}$_{\textrm{C}}$- and \textit{I}$_{\textrm{C}}$-bands,

\textit{b}, \textit{v}, \textit{r}, \textit{i} -- instrumental magnitudes in filters \textit{B}, \textit{V}, \textit{R}$_{\textrm{C}}$, \textit {I}$_{\textrm{C}}$, reduced to zenith distance z = 0, calculated as $-2.5\cdot \lg(N)$ $-\ \alpha\cdot X$, where $N$ is the number of counts (ADU) per second acquired in the 2.8 $e^{-}$/ADU gain mode, $\alpha$ is the extinction coefficient, $X$ is the air mass.

We built equations from measurements of 36 stars (in the range of colours not exceeding 0.6 mag) in the field NGC7654, which was observed at a zenith distance z $\sim$ 18$^{\circ}$ on September 22, 2020. The measured extinction coefficients on this night were:

\begin{gather*} 
\alpha_{B} = 0^{\textrm{m}}.50 \pm 0^{\textrm{m}}.030 \\ 
\alpha_{V} = 0^{\textrm{m}}.39 \pm 0^{\textrm{m}}.028\\
\alpha_{R_{\textrm{C}}} = 0^{\textrm{m}}.29 \pm 0^{\textrm{m}}.025\\
\alpha_{I_{\textrm{C}}} = 0^{\textrm{m}}.28 \pm 0^{\textrm{m}}.039
\end{gather*}

For our monitoring tasks, typical magnitudes of observed objects are 16 mag in the $V$-band. For 10 minutes of total exposure within a typical seeing of about 2$''$ at SAO, the accuracy for a star-like object is 0.005 mag. Providing the photometry of faint sources in a $V$-band on a single frame with an exposure time of 20 minutes for 22.5 mag we achieved $S/N \approx 4$ within a 1$''$.1 seeing.

\subsection{Polarimetry}\label{subsec10}

In the \texttt{MAGIC} device, the polarization analyzer is installed in a parallel beam. The design of the device involves the use of any type of polarization analyzer -- both a classic dichroic polaroid and birefringent prisms. At the moment, we use a double Wollaston prism for tasks of AGN polarimetry. The advantage of this analyzer is the ability to apply the \textit{one-shot polarimetry} approach when the number of images of the FoV sufficient to calculate the Stokes parameters is simultaneously registered at the detector in several angles of electric vector oscillation. This method minimizes the effect of atmospheric depolarization \citep[for more details see][]{AfAm}.

We use the quadrupole Wollaston prism, originally described in \cite{geyer}. The prism was produced by OPTEL\footnote{\url{https://optel.ru/}} and consists of two Wollaston calcite prisms glued together with a total size of 30$\times$30$\times$16 mm. The antireflection coating applied to the prism optics provides a high transmission of about 90\%, which leads to QE $\sim$ 45\% concerning the contribution of the device optics, CCD and telescope (Fig. \ref{fig:qe}). To avoid overlapping images in different polarization directions, the prism is used in conjunction with a square mask giving a square FoV in each direction of 6$'$.4$\times$6$'$.4.

\begin{figure}%[!tbp]
  \centering
  \begin{minipage}[b]{0.45\textwidth}
    \includegraphics[width=\textwidth]{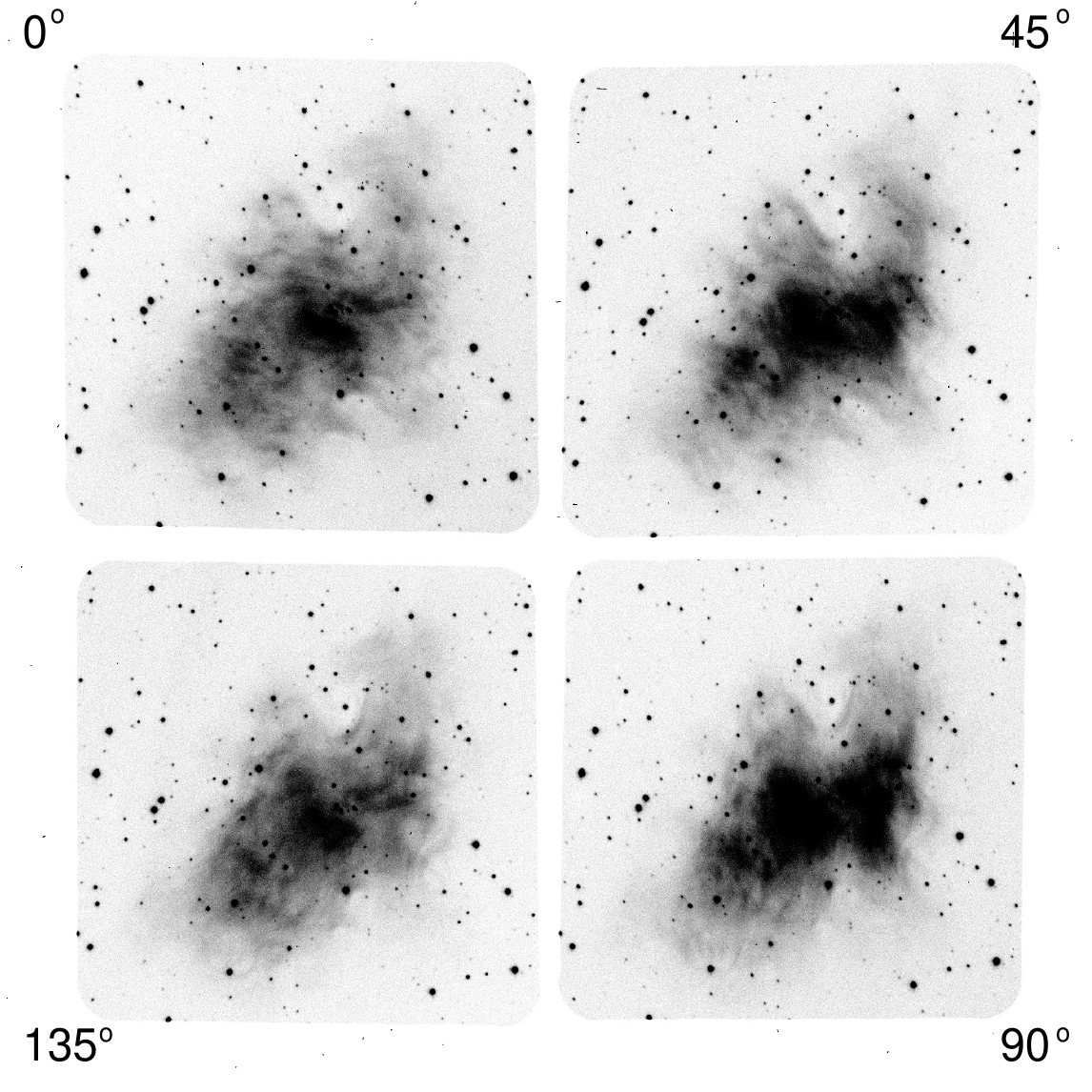}
    \caption{Observation of M1 in four directions of polarization (each FoV = 6$'$.4) with the quadrupole Wollaston prism in the SED600 filter ($t_{\rm exp} = 300$ s).}
    \label{fig:m1_raw}
  \end{minipage}
  \hfill
  \begin{minipage}[b]{0.48\textwidth}
    \includegraphics[width=\textwidth]{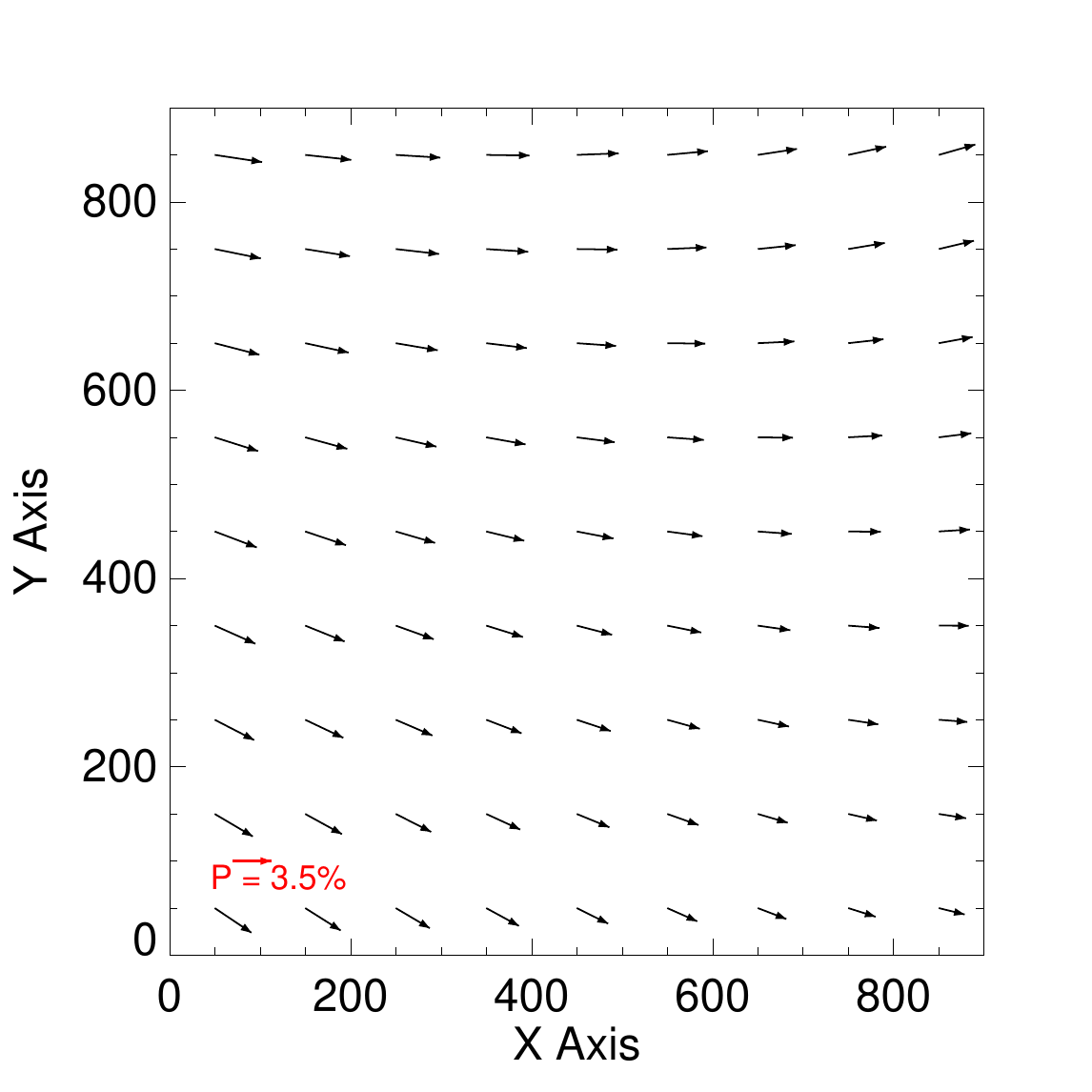}
    \caption{Instrumental polarization over the field inside the FoV of the quadrupole Wollaston prism. Coordinates in pixels are given along the X and Y axes, the coordinate grid is corrected for geometric distortions.}
    \label{fig:meas_ipol}
  \end{minipage}
\end{figure}

As an example, Fig.~\ref{fig:m1_raw} shows a frame of the M1 nebula, obtained with a Wollaston prism for 300 seconds of exposure in the SED600 filter. As can be seen, four directions of polarization are registered on the detector in the angles 0$^\circ$, 90$^\circ$, 45$^\circ$ and 135$^\circ$. This makes it possible to calculate three Stokes parameters $I$, $Q$, $U$, which describe the intensity and linear polarization of radiation, as follows:

\begin{gather*} 
I = I_0 + I_{90} + I_{45} + I_{135}, \\ 
\frac{Q}{I} = \frac{I_0 - I_{90} }{I_0 + I_{90} } , \\
\frac{U}{I} = \frac{I_{45} - I_{135} }{I_{45} + I_{135} } ,
\end{gather*}
where $I_0$, $I_{90}$, $I_{45}$, $I_{135}$ are the intensity in each direction, respectively. Further, for convenience, we will use the notation $Q \equiv Q/I$ and $U \equiv U/I$. The degree of polarization $P$ and the angle of the plane of polarization $\varphi$ are calculated by the formulas:
\begin{gather*}
{P} = \sqrt{Q^2 + U^2},\\
{\varphi} = \frac{1}{2} \arctan (Q/U).
\end{gather*}
Note that to rotate the Stokes parameters to the celestial plane, the Stokes vector should be multiplied by the rotation matrix of the $-$2$\cdot$PA angle, where PA is the instrument position angle.

\begin{figure*}
    \centering
    \includegraphics[scale=0.48,angle=90]{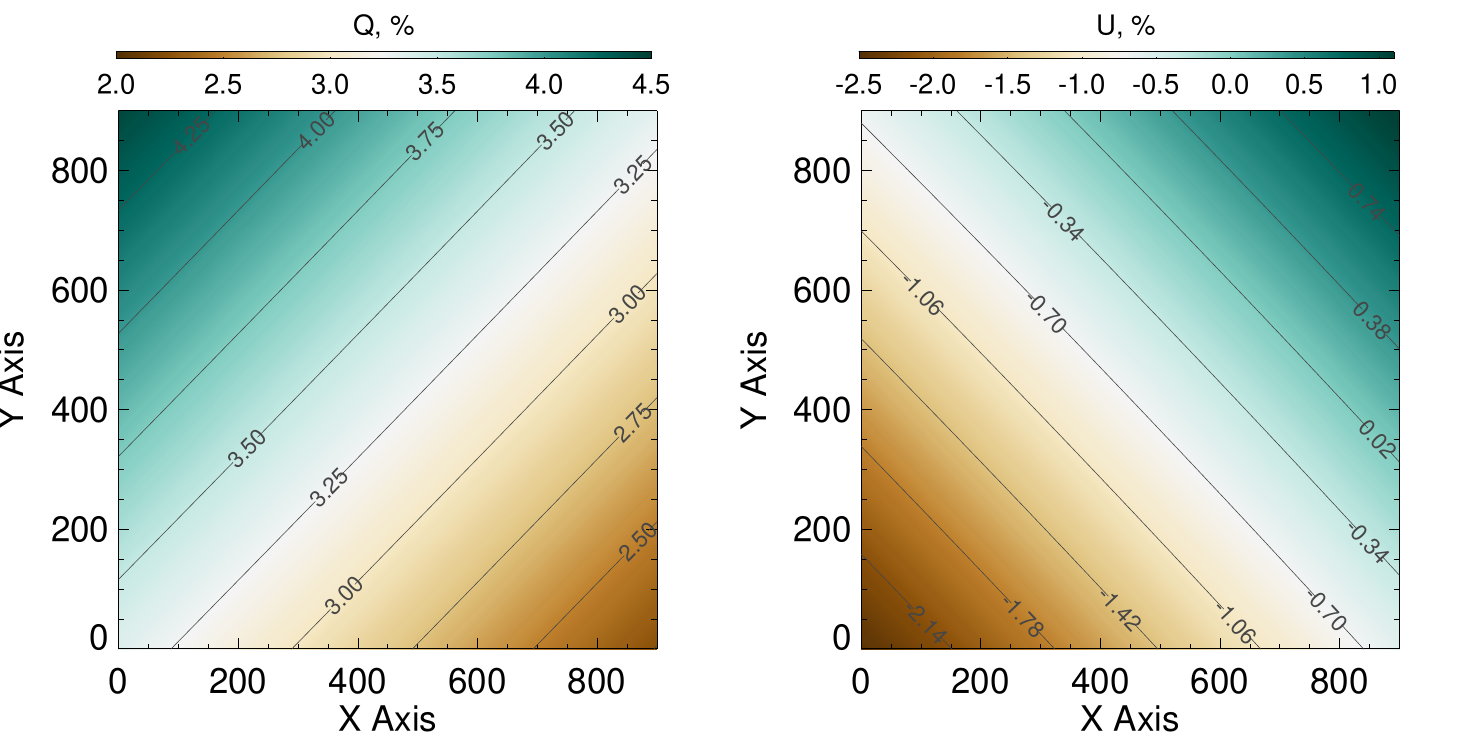}
    \caption{For the Stokes parameters $Q$ and $U$, smooth variations over the field inside the square mask are described.}
    \label{fig:ipols}
\end{figure*}

\begin{figure*}
    \centering
  \includegraphics[scale=0.48, angle=90]{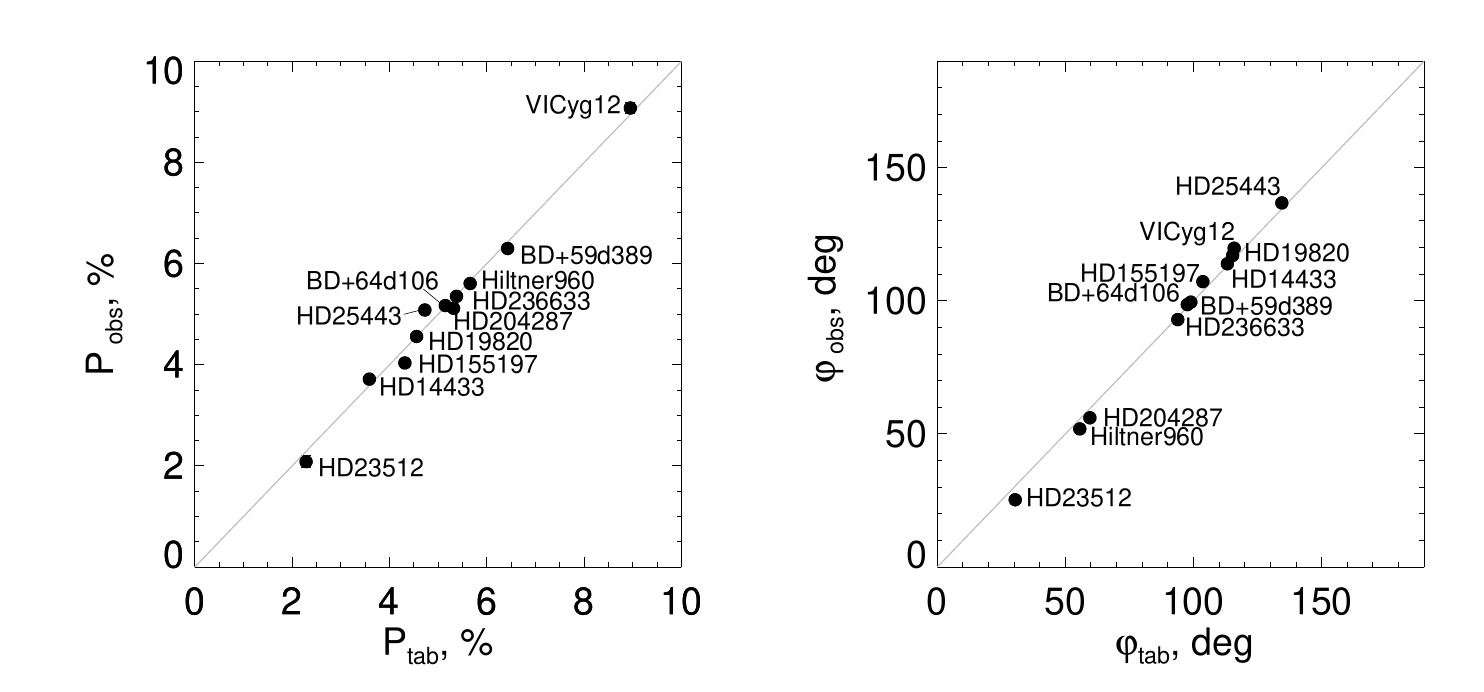}
    \caption{Comparison of the measured values of the degree of polarization $P_{\rm obs}$ (\textit{left}) and the polarization angle $\varphi_{\rm obs}$ (\textit{right}) with their reference values $P_{\rm tab}$ and $\varphi_{\rm tab}$.}
    \label{fig:ptab}
\end{figure*}

\begin{figure*}
    \centering
    \includegraphics[angle=90,scale=0.47]{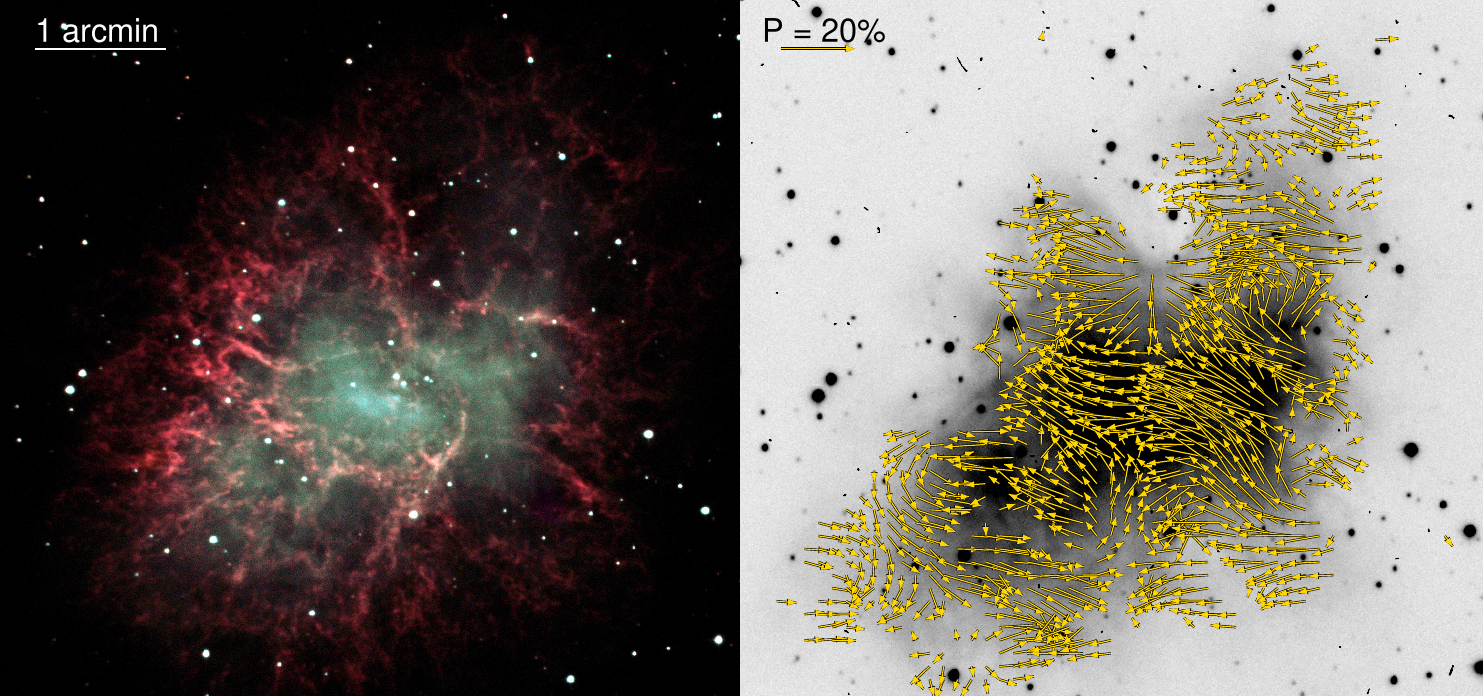}
    \caption{Results of observations of the M1 nebula: \textit{on the left}, a combined photometric image of the nebula in the $B$ (blue), $V$ (green), and SED650 (red) filters; \textit{on the right} is the polarization map of the nebula obtained with the Wollaston quadrupole prism in the SED600 filter.}
    \label{fig:M1_vector}
\end{figure*}

Due to the huge image separation, the prism used in \texttt{MAGIC} has its own dispersion, much larger than the more classic wedged version. Without the use of a filter in white light, the dispersion will decompose the star-like source image into a low-dispersion spectrum of $>$40$''$ in length. 
%(the plot of beam deflection by the prism dispersion is given in Appendix 1). 
The use of broadband filters, for example, the \textit{BVR}$_{\textrm{C}}$\textit{I}$_{\textrm{C}}$ system, with this prism is also not justified, since the distortions introduced by dispersion will be an order of magnitude greater than seeing. For this reason, observations with this quadrupole Wollaston prism are optimally carried out in medium-band filters.

Using the observations of unpolarized standard stars, we estimated the value of the instrumental polarization of the device within the FoV inside the mask. Repeated observations of zero standards at different positions in the field, as well as measuring the polarization of images forming by the 8-dots mask, which we use to correct geometric field distortions, we found that the changes of polarization are stable over time and have a smooth field dependence (Fig. \ref{fig:meas_ipol}). The average value of the degree of polarization $P$ introduced by the device is 3.5\% and varies over the field from 2.3\% to 4.5\%. The pattern and absolute values of the instrumental polarization do not change with the wavelength in the range 6000-7000 \AA\AA{}. Our laboratory tests of the optics and detector with other polarization analyzers introduced into the beam showed that the source of instrumental polarization is the prism.

We have described the  $Q$ and $U$ changes by 1st-order surfaces (Fig. \ref{fig:ipols}). After correcting observations of unpolarized stars for instrumental polarization using this model, the deviations of the parameters $Q$ and $U$ from zero were less than 0.05\%. Thus, the correction of instrumental polarization makes it possible to carry out high-precision polarimetric observations.

To determine the accuracy of the data obtained in the polarimetric mode, we observed a set of highly polarized standard stars. In Fig. \ref{fig:ptab} the dependence of the observed polarization degree $P$ and polarization angle $\varphi$ for a set of standard high polarization stars (after correction for instrumental effects) are plotted against their reference values. The deviations were $\Delta P = 0.18$\% and $\Delta \varphi = 3^\circ$. In general, according to our observations, for a star-like target up to 14 mag in medium-band filters with a seeing of 1$''$ for 20 minutes of total exposure, the polarization accuracy is better than 0.6\%.

The large field of view in the one-shot polarimetry mode is an important advantage for polarization observations of extended objects. An example of the results of such observations is shown in Fig.~\ref{fig:M1_vector}. For the Crab Nebula M1, a map of the change in the polarization of the continuum ('amorphous'{}) radiation was obtained, which makes it possible to compare the polarization characteristics of the nebula with its geometry. The measurement of the surface polarization was conducted for a methodical purpose and repeated the results obtained over the extensive history of Crab polarimetric studies initiated by \citet{baade} and subsequently analyzed by \citet{woltjer}. Our observations are in agreement with the surface polarization distribution, its degree, and orientation, as previously identified in earlier photographic studies \citep{baade,woltjer}, as well as in the initial CCD observations \citep{crabccd} with a large FoV similar to that of \MAGIC{}. These results are also consistent with \textit{HST} observations using a smaller FoV \citep{crabhst}.

\subsection{Long slit spectroscopy}\label{subsec11}

The spectral mode of the \texttt{MAGIC} device is implemented by introducing into the collimated beam (between the camera and the collimator) a direct vision grism VPHG600@500 (600 lines/mm, 500 nm -- central wavelength), as well as a slit into the converging beam in front of the collimator.
%, which occupies a position in turrets. 
The efficiency of the device in the spectral mode (telescope + optics + grating + CCD) is $\sim$16\% at maximum (Fig. \ref{fig:qe})\footnote{The efficiency here was also measured by on-sky standard stars. During the observations, the seeing was comparable to the slit width, and the slit losses of $\sim$80\% are taken into account. }.

The slit dimensions of 0.11 mm $\times$ 46 mm correspond to the angular dimensions 1$''.7$ $\times$ 12$'$ in the focal plane. The width of the projected monochromatic slit image onto the CCD plane is FWHM = 3.5 pix. We chose the slit sizes to achieve the best compromise between optimal CCD sampling, the required \textit{extragalactic}\footnote{Here~is~meant a compromise for studies of extragalactic objects between the spectral resolution for typical extragalactic tasks and denser concentration of light in a single CCD pixel.} spectral resolution, and minimizing light loss at the slit under average SAO weather conditions. In conjunction with the spectral grating, low-resolution spectra are obtained in the range 4000--7200 \AA\AA\ with reciprocal dispersion 2\AA/pix and spectral resolution $\delta\lambda \sim$ 7-8 \AA\ or in terms of ${R} =  \lambda$/$\delta\lambda \sim 1000$.

\begin{figure*}
    \centering
    \includegraphics[width=0.85\textwidth]{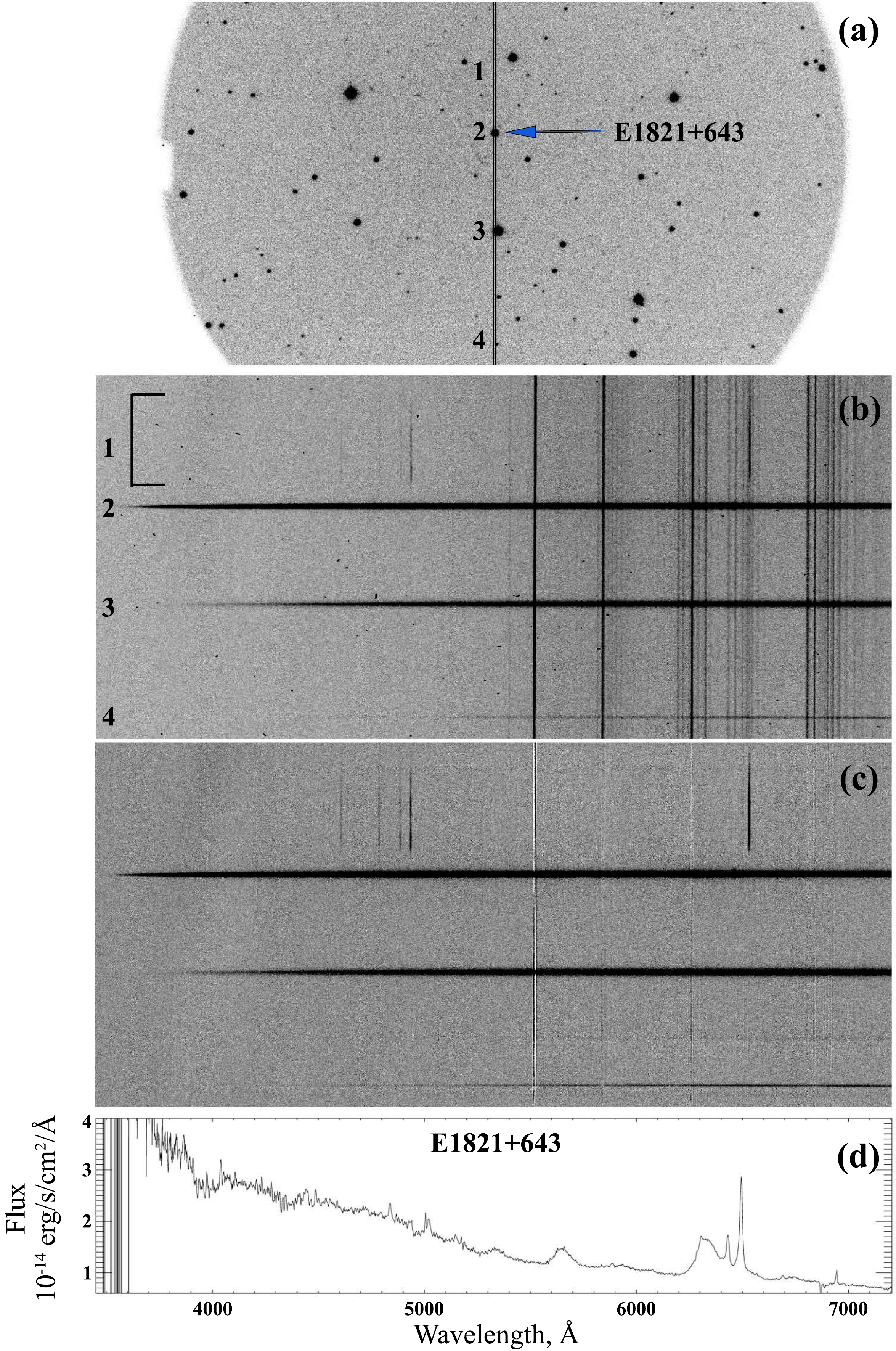}
    \caption{\MAGIC{} spectroscopy of the E1821+643 quasar: (a) a fragment of a direct image in the $R_{\textrm{C}}$ filter ($t_{\rm exp}$ = 10 sec) with the position of the spectrograph slit into which four objects fall, the arrow indicates the studied quasar; (b) -- single spectral frame ($t_{\rm exp}$ = 600 sec), contains traces of cosmic particles; (c) -- robustly averaged frame ($t_{\rm exp}$ = 8 $\times$ 600 sec) with geometric correction and subtracted night sky spectrum; (d) integrated spectrum in the wavelength scale of the quasar E1821+643. Marked in the figure: 1 -- planetary nebula PN K 1-16; 2 -- quasar  E1821+643; 3 -- star [SPB96] 1882; 4 -- field star.}
    \label{fig:spectra}
\end{figure*}

In Fig. \ref{fig:spectra} the sequence of obtaining observational material on the example of spectroscopy of type 1 AGN E1821+643 is demonstrated from setting the object onto the slit (in the direct image mode) to obtain the processed 1D spectrum. Observations are taken on September 21, 2020. It is interesting to note that in the presented frames, due to such a long slit, several objects are simultaneously observed, including the extended planetary nebula PN K 1-16 (indicated by number 1 in Fig. \ref{fig:spectra}). It is clear that the slit height of 12$'$ allows efficient spectroscopic observations of strongly extended objects, for example, comets. Such a long slit also simplifies sky subtraction when processing spectra.

At the moment, the development of a calibration module is underway to obtain auxiliary frames of a spectral flat field and a reference illumination of a He-Ne-Ar lamp for constructing a dispersion curve. However, only slight bending of the device (within $\pm$1 pix) makes it possible to use an auxiliary appliance installed on the inside of the telescope dome (see Fig. \ref{fig:MAGIC_flat}, on the right) to obtain calibration frames, which gives Lambertian scattering under illumination lamp.

\section{Conclusions}\label{sec5}

In 2020, the \MAGIC{} multi-mode focal reducer for the 1-m Zeiss-1000 telescope of the SAO RAS was designed, manufactured and put into operation. The device effectively solved the problem of oversampling in the Cassegrain focus, making the optical system faster (from $F/13.3$ to $F/6.1$) and more effective for the study of faint and/or extended objects. The optics of the device constrain an $\sim0''.3$ image of a point source and has an integral transmission QE $\sim 50\%$. 
%The focal reducer is used in conjunction with the Andor iKon-L 936 CCD system with an BEX2-DD type 2048 $\times$ 2048 pix E2V CCD42-40 with quantum efficiency $>90$\% in the range of 400-850 nm and not less than 40\% in the range of 340-990 nm. Taking into account the absence of fringes (interference pattern), the focal reducer together with this CCD makes it possible to observe faint sources in a wide range of wavelengths up to almost $1\mu$m. 
The ability to observe and quickly switch between observation modes allows one to respond flexibly to the weather conditions changes during the night, as well as to comprehensively explore astrophysical objects. Currently, three observation modes are implemented in the \MAGIC{} device.

\begin{itemize}
    \item Direct images could be taken in the Johnson-Cousins photometric system and in the medium-band interference filters. The photometry FoV is $\sim$12$'$ with a scale of 0$''$.45/pix. The filters are set in 2 wheels each of 9 positions. For 10 minutes of total exposure within a typical seeing of about 2$''$ at SAO, the accuracy for a star-like object of 16 mag in $V$-band is 0.005 mag. The limited magnitude in $V$-band ($S/N \approx 4$) is 22.5 mag within a 1$''$.1 seeing and 20 minutes exposure.
    
    \item Image-polarimetry mode provides measurements of intensity and linear polarization in $6'.4 \times 6'.4$ FoV. The introduced instrumental polarization varies over the field and could be compensated by the calculated smooth model. For a star-like target up to 14 mag in medium-band filters with a seeing of 1$''$ for 20 minutes of total exposure, the accuracy of the intensity measurement is better than 0.01 mag and the polarization accuracy is better than 0.6\%. 
    
    \item In long-slit spectroscopy the combination of 1$''.7$ $\times$ 12$'$ slit and volume phase holographic disperser VPHG600@500 is used. Low-resolution spectra are obtained in the range 4000--7200 \AA\AA\ with reciprocal dispersion 2\AA/pix and spectral resolution $\delta\lambda \sim$ 7-8\AA{}.
\end{itemize}

To use the \MAGIC{} device on the 1-m Zeiss-1000 telescope, the optomechanical scheme of the telescope was upgraded. The modernization of baffles made it possible to minimize parasitic rays in the telescope tube, correcting additive noises that occurred in observations. The installation of additional modules -- a rotator and an offset guide -- helps to solve the problem of accurate telescope guidance and instrument orientation.

It is important to note that exactly the given optical scheme and design can be used to create universal devices for a wide class of small Cassegrain telescopes with a large focal ratio ($\lesssim F/8$) and a large aberration-free FoV. A specific implementation of the \MAGIC{} device is a fairly universal solution to reduce the relative focus of the system for a large number of both already built Zeiss-type telescopes and new ones. The realizable efficiency of \MAGIC{} makes it possible to carry out joint monitoring campaigns in conjunction with other focal reducers [see, e.g., results of \MAGIC{} observations in \citep{polrev} obtained together with AFOSC of the 1.82-m Copernico telescope of Asiago-Cima Ekar observatory and FoReRo-2 of the 2-m telescope of Rozhen National Astronomical Observatory], as well as to carry out observations applying the original methodical approaches [see, e.g., the Stokes polarimetry of blazars with quadruple Wollaston prism in two-band filter \citep{camel}].

\section*{Acknowledgements}

\MAGIC{} was the last of many astronomical devices created by Viktor Leonidovich Afanasiev (1947 -- 2020). We will remember him as a brilliant practising astronomer who deeply understands the experiment -- from the formulation of scientific issues, the device creation and development of observational techniques to the obtaining of observational data and its competent interpretation. He loved science and was an ideological inspirer. His contribution to the development of our observatory is invaluable.

We are grateful to E.I. Perepelitsyn for the manufacture of optics for the device. The mechanical and optical parts of \MAGIC{}, as well as parts for the modernization of the telescope units, were produced at the SAO breadboard workshops. We also thank the engineers of the 1-m Zeiss-1000 telescope led by V.V. Komarov for constant assistance in the work with the telescope. We thank Dr. Imre Barna B\'ir\'o for helpful discussions and advice on baffles. We express our gratitude to A.V. Moiseev for providing valuable methodological guidance throughout the study of the device. Also, we appreciate the constructive comments provided by the reviewers, which significantly enhanced the quality of this paper.

This work was supported by the Russian Scientific Foundation (grant no. 20-12-00030 "Investigation of geometry and kinematics of ionized gas in active galactic nuclei by polarimetry methods"). 
Observations with the SAO RAS telescopes are supported by the Ministry of Science and Higher Education of the Russian Federation.

\section*{Data Availability}

The data underlying this article will be shared on reasonable request to the corresponding author.



\bibliographystyle{mnras}
\bibliography{example} % if your bibtex file is called example.bib

%\begin{thebibliography}{99}
%\end{thebibliography}

%%%%%%%%%%%%%%%%%%%%%%%%%%%%%%%%%%%%%%%%%%%%%%%%%%

%%%%%%%%%%%%%%%%% APPENDICES %%%%%%%%%%%%%%%%%%%%%

\appendix

\section{CCD inhomogeneities in near-infrared light}\label{secA1}

\begin{figure*}
    \centering
    \includegraphics[width=1.0\textwidth]{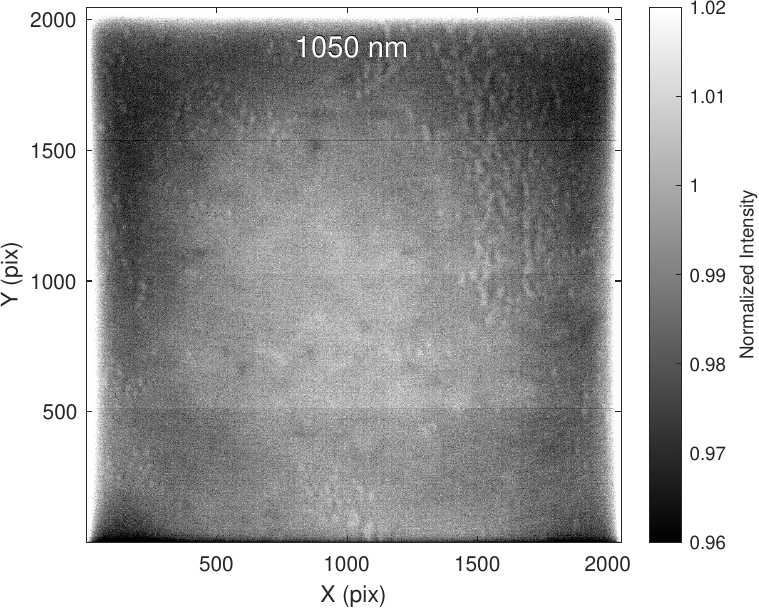}
    \includegraphics[trim=20 0 0 0, width=0.85\textwidth]{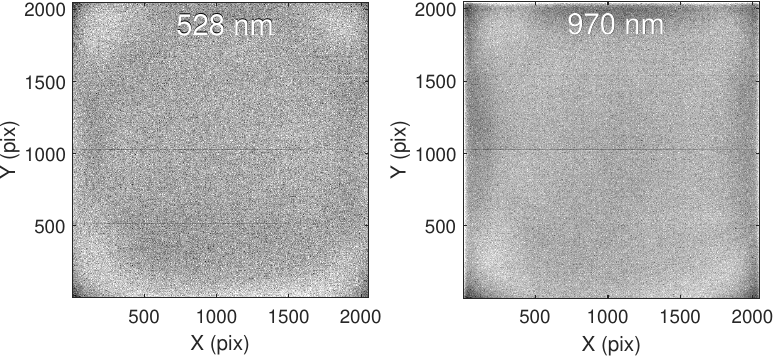}
    \caption{The frames (2048 $\times$ 2048 pix) normalized to the average intensity of exposure to illumination at wavelengths of 1050, 528 and 970 nanometers.}
    \label{fig:micron}
\end{figure*}

We performed a check-up of the Andor iKon-L 936 CCD system with a BEX2-DD type E2V CCD42-40 in the laboratory to control the interference patterns that usually appeared while using thin CCD. 
We noticed inhomogeneities of the order of 2\% only when exposed to illumination with $\lambda$\ = 1050 nm, while the noise did not correspond to the classical interference pattern. An example of a frame normalized to average intensity is shown in Fig.~\ref{fig:micron} (top panel). These were rounded inhomogeneities, the nature of which can be different (from manifestations of the optic at $\lambda > 1$\ $\mu$m to thermal noise from electrodes).
No noise was detected at shorter wavelengths. For comparison, Fig.~\ref{fig:micron} (bottom panel) shows frames illuminated by LEDs with wavelengths of 528 nm and 970 nm (with FWHM of the order of 10 nm).
Thus, this CCD allows useful research in the red part of the spectrum, without complicating observations with various interferences like fringes, at high sensitivity.

%%%%%%%%%%%%%%%%%%%%%%%%%%%%%%%%%%%%%%%%%%%%%%%%%%

% Don't change these lines
\bsp	% typesetting comment
\label{lastpage}
\end{document}